\documentclass[fleqn,usenatbib]{mnras}

\usepackage{latexsym,mathrsfs,amssymb,bm}
\usepackage[tbtags]{amsmath}
\usepackage[T1]{fontenc}
\usepackage{ae,aecompl,times}
\usepackage{epsfig, ulem}
\usepackage[dvipsnames]{xcolor}
\usepackage{multirow}
\usepackage{tikz,array,color,float}
\usepackage[utf8]{inputenc}
\usepackage[T1]{fontenc}
\usepackage{graphicx}
\usepackage{caption}
\usepackage{multirow}

\title[Ray-tracing the MTNG simulations]{Ray-tracing vs. Born approximation in full-sky weak lensing simulations of the MillenniumTNG project}

\author[F. Ferlito et al.]{%
\parbox{0.98\textwidth}{
Fulvio Ferlito$^{1}$\thanks{E-mail: \href{mailto:ferlito@mpa-garching.mpg.de}{ferlito@mpa-garching.mpg.de}},
Christopher T. Davies$^{2}$,
Volker Springel$^{1}$,
Martin Reinecke$^{1}$,
Alessandro Greco$^{3}$, \\
Ana Maria Delgado$^{6}$,
Simon D. M. White$^{1}$,
C\'esar Hern\'andez-Aguayo$^{1,4}$,
Sownak Bose$^{5}$ \\
and Lars Hernquist$^{6}$
}
\vspace*{0.2cm}\\%
$^{1}$Max-Planck-Institut f\"ur Astrophysik, Karl-Schwarzschild-Str. 1, D-85748, Garching, Germany\\%
$^{2}$Faculty of Physics, Ludwig-Maximilians-Universit\"at, Scheinerstr. 1, 81679 Munich, Germany\\%
$^{3}$Department of Astronomy, University of Florida, 211 Bryant Space Science Center, Gainesville, FL 32611, USA\\%
$^{4}$Excellence Cluster ORIGINS, Boltzmannstrasse 2, D-85748 Garching, Germany\\%
$^{5}$Institute for Computational Cosmology, Department of Physics, Durham University, South Road, Durham, DH1 3LE, UK\\%
$^{6}$Center for Astrophysics | Harvard $\&$ Smithsonian, 60 Garden St, Cambridge, MA 02138, USA\\%
}

\date{Accepted XXX. Received YYY; in original form ZZZ}
\pubyear{2024}

\begin{document}

\label{firstpage}

\pagerange{\pageref{firstpage}--\pageref{lastpage}}

\maketitle

\begin{abstract}
Weak gravitational lensing is a powerful tool for precision tests of cosmology. As the expected deflection angles are small, predictions based on non-linear N-body simulations are commonly computed with the Born approximation. Here we examine this assumption using {\small DORIAN}, a newly developed full-sky ray-tracing scheme applied to high-resolution mass-shell outputs of the two largest simulations in the MillenniumTNG suite, each with a 3000 Mpc box containing almost 1.1 trillion cold dark matter particles in addition to 16.7 billion particles representing massive neutrinos. We examine simple two-point statistics like the angular power spectrum of the convergence field, as well as statistics sensitive to higher order correlations such as peak and minimum statistics, void statistics, and Minkowski functionals of the convergence maps. Overall, we find only small differences between the Born approximation and a full ray-tracing treatment. While these are negligibly small at power-spectrum level, some higher order statistics show more sizable effects; ray-tracing is necessary to achieve percent level precision. At the resolution reached here, full-sky maps with 0.8 billion pixels and an angular resolution of 0.43 arcmin, we find that interpolation accuracy can introduce appreciable errors in ray-tracing results. We therefore implemented an interpolation method based on nonuniform fast Fourier transforms (NUFFT) along with more traditional methods. Bilinear interpolation introduces significant smoothing, while nearest grid point sampling agrees well with NUFFT, at least for our fiducial source redshift, $z_s=1.0$, and for the 1 arcmin smoothing we use for higher-order statistics.
\end{abstract}

\begin{keywords}
gravitational lensing: weak -- methods: numerical -- large-scale structure of the Universe
\end{keywords}

\section{Introduction}
\label{sec:intro}

Distant galaxies appear weakly sheared to the observer due to the differential deflection of light by the foreground matter distribution. This effect, known as weak gravitational lensing \citep[hereafter WL; for reviews see, e.g.,][]{Bartelmann2001,  Hoekstra2008, Kilbinger2015, Mandelbaum2018} is among the most relevant physical processes that help us to investigate and understand the cosmic matter distribution. WL surveys of the previous decade, like the stage-III projects HSC \citep{Aihara2022}, DES \citep{Abbott2022} and KiDS \citep{Heymans2021}, have already produced important constraints on the cosmological parameters, suggesting a tension with the value of $S_8$ inferred from microwave background observations \citep{Hildebrandt2016}, as well as insights into the nature of dark matter and dark energy. The next generation of WL surveys (named stage-IV), including Rubin \citep{LSST}, Euclid \citep{Amendola_2018}, and Roman \citep{Roman}, will provide a substantial increase both in sky coverage and in angular resolution, thereby increasing constraining power substantially. To fully exploit these rich and complex data, it is crucial to develop high-fidelity modeling techniques for WL observables, based on accurate numerical simulations.

At present, the most popular WL cosmological probe is the so-called ``3×2pt'' statistic ~\citep[see e.g.][]{Abbott2018}. This combines three two-point correlation functions: galaxy clustering, WL galaxy shear, and the galaxy-shear cross-correlation. Additionally, higher-order WL statistics have gained popularity and shown their efficacy for extracting complementary cosmological information. Some examples include: one-point PDF~\citep[][]{Liu2019, Boyle2021}, counts of peaks and/or minima~\citep[][]{Martinet2018, Coulton2020, Davies2022, Marques2024}, Minkowski functionals~\citep[][]{Grewal2022}, voids~\citep[][]{Davies2021, Boschetti2023}, bispectrum~\citep[][]{Rizzato2019}, trispectrum~\citep[][]{Munshi2022}, aperture mass statistic~\citep[][]{Schmalzing1998, Martinet2021}, three-point correlation function~\citep[][]{Takada2003}, and integrated three-point correlation function~\citep[][]{Halder2023}.

Accurately computing the above statistics, either numerically or analytically, is a significant challenge, as many approximations need to be employed. A non-exhaustive list of assumptions often made in numerical WL experiments includes: the Limber and flat-sky approximations~\citep[][]{Lemos2017}; the projection of 3D mass distributions into infinitely thin planes, i.e. the thin lens approximation~\citep[][]{Frittelli2011, ParsiMood2013, Zhou2024}; neglecting higher-order image distortions beyond convergence and shear, e.g.~flexions \citep{Schneider2008}; the use of DM-only simulations rather than including full baryonic physics \citep[][]{Semboloni2011,Yang2013,Osato2021,Broxterman2024,Ferlito2023}; and simplifying assumptions for the sample redshift distribution $n(s)$~\citep[][]{Zhang2023}.

Finally, one of the most common approximations used in calculating WL from simulations is the Born approximation. This assumes that the perturbations to the light path induced by gravitational lensing are negligible, so that it can be well approximated by an undeflected straight line. Carrying out a WL simulation in the Born approximation requires significantly less memory and computational effort than tracing the paths of rays explicitly~\citep[see e.g.][]{Petri2017}, making it particularly attractive. The accuracy of the Born approximation in WL simulations has previously been studied for square maps~\citep[][]{Petri2017}, and, in the case of lensing of the cosmic microwave background (CMB), also for full-sky maps~\citep[][]{Fabbian2018}. These studies have shown its impact to be effectively negligible at the level of the power spectrum~\citep[see also e.g.][]{Fabbian2019, Hilbert2020}. Nevertheless, it has also been found to have a non-negligible impact on higher moments of the convergence PDF; i.e.~the skewness and kurtosis~\citep[][]{Petri2017,Fabbian2018,Barthelemy2020}.

In this work, we revisit this question and develop a full-sky ray-tracing scheme that works on the lightcone mass-shell outputs produced by the {\small GADGET-4} code~\citep{Springel2021} for the MillenniumTNG simulation suite. This allows us to explicitly test the Born approximation, not only on the power spectrum and PDF, but also on a set of popular higher-order statistics: counts of peaks and minima, the abundance and profiles of voids, and Minkowski functionals.

This paper is organized as follows. In Section~\ref{sec:theory}, we give a theoretical overview of weak gravitational lensing. In Section~\ref{sec:methods}, we describe the numerical simulations and methods employed in this work. We first introduce the MillenniumTNG simulations and their mass-shell output (Sec.~\ref{subsec:mtng}), we discuss our implementation of a ray-tracing scheme (Sec.~\ref{subsec:raytracing}), and we then focus on the spherical harmonics relations (Sec.~\ref{subsec:sph}) and the interpolation schemes (Sec.~\ref{subsec:interpolation}) used. At the end of the section, we provide additional details relevant to the computation of observables (Sec.~\ref{subsec:observables}). In Section~\ref{sec:results}, after a qualitative discussion of our full-sky maps, we show the impact of ray-tracing on the following statistics: the angular power spectrum (Sec.~\ref{subsec:powerspectrum});  the PDF and counts of peaks and minima of the convergence (Sec.~\ref{subsec:pdf}, \ref{subsec:peaks_minima}); void statistics (Sec.~\ref{subsec:voids}); and Minkowski functionals (Sec.~\ref{subsec:minkowski}). Finally, in Section~\ref{sec:conclusions} we summarize our conclusions, and in~Appendix~\ref{appendix:interpolation} we discuss in further detail the effects caused by bilinear interpolation.


\section{Theoretical background}
\label{sec:theory}

In this section, we first introduce the key quantities of weak lensing. Then we present expressions for these quantities in the spherical harmonics domain, as this will be required to describe the implementation of ray-tracing in subsection~\ref{subsec:raytracing}.

\subsection{Weak gravitational lensing formalism}
\label{subsec:wlform}

We adopt the Friedmann-Lemaître-Robertson-Walker cosmology with small scalar inhomogeneous perturbations that can be expressed in terms of the Newtonian gravitational potential $\Phi$ \citep[see, e.g.,][]{Kaiser1998} to describe our lensing system. We use the lens equation from geometric optics to relate the observed angular position $\bm{\theta}$ to the true angular position $\bm{\beta}$ of a ray of light that, encountering a lens (or a system of lenses), is deflected by an angle $\bm{\alpha}$: 
\begin{equation}
\label{eq:lens_eq}
\bm{\beta} = \bm{\theta} - \bm{\alpha} \, .
\end{equation}

In the context of gravitational lensing, matter acts as a lens, thus leading to the gravitational lens equation, which tells us that the observed position of a light ray starting from redshift $z_{\rm s}$ will depend on the surrounding matter field during its travel to the observer. The deflection angle is then:
\begin{equation}
\label{eq:grav_lens_eq}
\bm{\beta}(\bm{\theta}, z_{\rm s}) = \bm{\theta} - \frac{2}{{\rm c}^2} \int^{\chi_{\rm s}}_0 {\rm d} \chi_{\rm d} \frac{f_{\rm ds}}{f_{\rm d} f_{\rm s}} \nabla_{\bm{\beta}} \Phi (\bm{\beta}(\bm{\theta}, \chi_{\rm d}), \chi_{\rm d}, z_{\rm d}) \, ,
\end{equation}
where we have introduced the speed of light $c$, the angular gradient $\nabla_{\bm{\beta}}$, the comoving line-of-sight distance $\chi$, and the comoving angular diameter distance $f_{K}(\chi)$. The subscripts ``s'' and ``d'' refer, respectively, to the source and the lens; hence the geometric factors are $f_{\rm ds} = f_{K}(\chi_{\rm s} - \chi_{\rm d})$, $f_{\rm d} = f_{K}(\chi_{\rm d})$ and $f_{\rm s} = f_{K}(\chi_{\rm s})$.

The distortion of an image, formed by the ray at $\bm{\theta}$ and the ones nearby, can be described by the distortion matrix $A$, obtained by differentiating the previous equation with respect to $\bm{\theta}$:
\begin{equation}
\label{eq:dist_matrix_def}
\begin{aligned}
A_{ij}  & \equiv \frac{\partial \beta_i (\bm{\theta}, z_{\rm s})}{\partial \theta_j} = \delta_{ij} - \frac{2}{{\rm c}^2} \int^{\chi_{\rm s}}_0 {\rm d} \chi_{\rm d} \frac{f_{\rm ds}}{f_{\rm d} f_{\rm s}} \\ 
&\times \frac{\partial^2 \Phi (\bm{\beta}(\bm{\theta}, \chi_{\rm d}), \chi_{\rm d})}{\partial \beta_i \partial\beta_k} \frac{\partial \beta_k \bm{\beta}(\bm{\theta}, \chi_{\rm d})}{\partial \theta_j} \, ,
\end{aligned}
\end{equation}
where $\delta_{ij}$ is the Kronecker delta. This matrix can be decomposed as follows: 
\begin{equation}
\label{eq:dist_matrix_comp}
\begin{aligned}
A &\equiv
\begin{pmatrix}
\phantom{-}\cos{\omega}  & \sin{\omega} \\
-\sin{\omega} & \cos{\omega}
\end{pmatrix}
\begin{pmatrix}
1 - \kappa - \gamma_1 & -\gamma_2  \\
-\gamma_2             & 1 - \kappa + \gamma_1
\end{pmatrix}
\\
&\approx
\begin{pmatrix}
1 - \kappa - \gamma_1 & -\gamma_2 + \omega \\
-\gamma_2 - \omega    & 1 - \kappa + \gamma_1
\end{pmatrix} \, ,
\end{aligned}
\end{equation}
where we have introduced three fundamental WL quantities: the convergence $\kappa$, the rotation $\omega$, and the shear $\gamma = \gamma_1 + {\rm i}\gamma_2$. In the WL regime, the distortion of images is small; i.e. the distortion matrix is close to the identity matrix. In such a regime, the rotation of images is tiny, justifying the approximation in the above equation.

A common way to approach the integral in equation~(\ref{eq:dist_matrix_def}), known as the Born approximation, consists of integrating along an unperturbed straight light path; i.e. directly over $\theta$ instead of $\beta$. This significantly simplifies the equation, leading to:
\begin{equation}
\label{eq:dist_matrix_born}
\frac{\partial \beta_i (\bm{\theta}, z_{\rm s})}{\partial \theta_j} = \delta_{ij} - \frac{2}{{\rm c}^2} \int^{\chi_{\rm s}}_0 {\rm d} \chi_{\rm d} \frac{f_{\rm ds}}{f_{\rm d} f_{\rm s}} \frac{\partial^2 \Phi (\bm{\theta}, \chi_{\rm d})}{\partial \theta_i \partial\theta_j} \, .
\end{equation}
This approximation, valid when the light rays experience little deflection, makes it possible to directly relate the convergence to the matter density contrast $\delta_{\rm m}$:
\begin{equation}
\label{eq:kappa_born}
\begin{split}
\kappa_{\rm{born}}&(\bm{\theta}, z_{\rm s})=\\
&= \int^{\chi_{\rm s}}_0 {\rm d} \chi_{\rm d} \, \frac{3 H_0^2 \Omega_m}{2 {\rm c}^2} (1+z_{\rm d}) \frac{f_{\rm d} f_{\rm ds}}{f_{\rm s}} \, \delta_{\rm m} (\bm{\theta}, \chi_{\rm d}, z_{\rm d})
 \, .
 \end{split}
\end{equation}
Here we have neglected boundary terms at the observer and source, and assumed that the Universe is in a matter-dominated epoch. 

The Born approximation is obtained by expanding equation~(\ref{eq:dist_matrix_def}) to linear order in terms of $\Phi$. By expanding to the quadratic order, one would obtain the following two additional terms:
\begin{equation}
\label{eq:kappa_quadratic}
\kappa = \kappa_{\rm{born}} + \kappa_{\rm{ll}} + \kappa_{\rm{geo}} + O(\Phi^3) .
\end{equation}
We refer the reader to equations~(9) and (10) of \citet{Petri2017} for the explicit expressions of $\kappa_{\rm{ll}}$ and $\kappa_{\rm{geo}}$, respectively. In the following, we briefly describe the physical meaning of these two terms. The first post-Born term, $\kappa_{\rm{ll}}$, accounts for non–local couplings between lenses at the quadratic level; in other words, it considers that light is not perturbed independently by the series of deflectors along the line of sight, but rather that the deformation from background lenses is progressively distorted by the foreground lenses. This term is the lowest-order one to introduce a non-zero rotation. The second post-Born term, $\kappa_{\rm{geo}}$, accounts for the actual bending of light rays by integrating the matter density contrast along the corrected path at the lowest order in the geodesic deﬂections, as opposed to a straight trajectory. For this reason, it is sometimes called the {\it Born correction} \citep[see e.g.][]{Cooray2002}.


\section{Methods}
\label{sec:methods}

\subsection{Simulations}
\label{subsec:mtng}

The simulations used in this work are a subset of the MillenniumTNG (MTNG) project, a recent simulation suite consisting both of large dark matter only simulations (some additionally with massive neutrinos), as well as matching fully hydrodynamical simulations in comparatively large volumes. We refer the reader to \citet{Aguayo2023} for an overview of the simulation suite, and to \citet{Pakmor2023} for details regarding the full-hydro run. First MTNG results on intrinsic galaxy alignments can be found in \cite{Delgado2023}, on the clustering of galaxies in \citet{Barrera2023, Bose2023}, on improving the accuracy of the halo occupation distribution (HOD) formalism in \citet{Hadzhiyska2023a, Hadzhiyska2023b}, on the high-redshift galaxy population in \citet{Kannan2023}, and on cosmological parameter inferences in \citet{Contreras2023}. In order to avoid repetitions of the simulation volume along the line of sight and to maximize statistical robustness, we employ the simulations of the MTNG suite which feature the biggest box size together with the highest particle number. These are the A- and B-realizations of the N-body cosmological run MTNG3000-DM-0.1$\nu$, which feature dark matter (DM) and massive neutrinos with a summed mass of $\Sigma m_\nu = 100\,{\rm meV}$. Both simulations were performed with the {\small GADGET-4} code \citep{Springel2021} and are characterized by a periodic box size of side length $2.04\,h^{-1}{\rm Gpc} = 3\,{\rm Gpc}$, and a number of particles used for cold dark matter and neutrinos equal to $N_{\rm cdm} = 10240^3$ and $N_{\nu}= 2560^3$, respectively. 

The gravitational softening for the dark matter particles has been set to $\epsilon_{\rm cdm} = 4\,h^{-1}\,{\rm kpc}$, corresponding roughly to $1/50$-th of the mean inter-particle separation. The effect of massive neutrinos is implemented via the $\delta f$ method proposed by \citet{Elbers:2020lbn}. The reader is referred to Hernández-Aguayo et al.~(2024, in prep) for further details on our implementation of neutrino physics. The simulations used the following cosmological parameters: $\Omega_{\rm m} = \Omega_{\rm cdm} + \Omega_{\rm b} = 0.3037$, $\Omega_{\rm b} = 0.0487$, $\Omega_\Lambda = 0.6939$, $h = 0.68$, $\sigma_8 = 0.804$ and $n_s=0.9667$. The initial conditions, set at $z=63$, are generated via second-order Lagrangian perturbation theory with an updated version of the {\small N-GENIC} code as part of {\small GADGET-4}. The initial conditions of the two runs employ the ﬁxed-and-paired technique for cosmic variance suppression proposed by \citet{Angulo2016}.

In continuation of \citet{Ferlito2023}, the main simulation product of interest for WL applications is the ``mass-shell'' output: a collection of concentric HEALPix maps \citep{Gorsky2005} produced during the simulation on the fly in an onion-like fashion with fixed comoving thickness. For each shell, every pixel stores the cumulative mass of particles intersecting the time-evolving hypersurface of the lightcone of a putative observer. In the present work, we use a HEALPix parameter of $N_{\rm side}=8192$, corresponding to $8.053 \times 10^8$ pixels and an angular resolution of 0.43 arcmin.

\subsection{Implementation of ray-tracing}
\label{subsec:raytracing}

Our ray-tracing implementation, dubbed {\small DORIAN}\footnote{Acronym for "Deflection Of Rays In Astrophysical Numerical simulations".}, is a \texttt{Python} code based on the multiple-lens-plane approximation~\citep[e.g.][]{Blandford1986,Schneider1992,Jain2000}, which has been adopted in a number of codes~\citep[see e.g.][]{Hilbert2009, Becker2013, Petri2016, Fabbian2018}. In our setup, each mass-shell (as described in the previous section), here labeled with an index $k$, constitutes a thin lens. The surface mass density distribution of the lens is given by
\begin{equation}
\label{eq:surfmassdens}
\Sigma^{(k)}(\bm{\beta}) = \frac{M(\bm{\beta})}{A_{\mathrm{pix}}} \, .
\end{equation}

This is obtained by dividing the mass assigned to each pixel by the pixel area $A_{\rm pix} = 4 \pi / N_{\rm pix}$ (in steradians). Using equation~(\ref{eq:kappa_born}), we can compute an approximation of the convergence at the $k^{\rm{th}}$ lens plane as:
\begin{equation}
\label{eq:kappa_plane}
\kappa^{(k)}(\bm{\beta}) = \frac{4 \pi \mathrm{G} }{\mathrm{c}^2}  \frac{(1+z^{(k)}_{\rm d})}{f^{(k)}_{K}} \left[ \Sigma^{(k)}(\bm{\theta}) - \bar\Sigma^{(k)} \right] \, ,
\end{equation}
where ${f^{(k)}}_K = f_{K}(\chi_{(k)})$. We introduce the lensing potential $\psi$ as the 2-D projection of the Newtonian gravitational potential onto the $k^{\rm{th}}$ lens surface:
\begin{equation}
\label{eq:lensing_potential}
\psi^{(k)}(\bm{\beta}) = \frac{2}{c^2 f^{(k)}_K} \int_{\chi_k - \Delta\chi/2}^{\chi_k + \Delta\chi/2} \, \Phi(\bm{\beta}) \rm d \chi,
\end{equation}
where $\Delta\chi$ is the comoving thickness of each shell. The above quantity is related to $\kappa^{(k)}$ through the Poisson equation:
\begin{equation}
\label{eq:potential_k_plane}
\nabla_{\bm{\beta}}^2 \psi^{(k)}(\bm{\beta}) = 2 \kappa^{(k)}(\bm{\beta}) \, ,
\end{equation}
where $\nabla_{\bm{\beta}}^2$ is the Laplacian operator. The multiple lens plane approximation neglects modes along the line of sight larger than the shell thickness, and therefore is expected to become unreliable as the lens planes become too thin \citep{Das2008}. This motivates us to set the shell thickness to $\Delta \chi = 100 \,h^{-1}{\rm Mpc}$ in the present work \citep[see also][]{ZorrillaMatilla2020}. We refer the reader to section 2.2 of \citet{Becker2013} and appendix B of \citet{Takahashi2017} for a more detailed discussion. 

By taking the first derivative of the lensing potential, one obtains the deflection angle at the $k^{\rm{th}}$ lens plane:  
\begin{equation}
\label{eq:deflection_k_plane}
\bm{\alpha}^{(k)}(\bm{\beta}) = \frac{\partial \psi^{(k)}(\bm{\beta})}{\partial \beta_i}  \, .
\end{equation}
The second derivatives of the lensing potential give the shear matrix $U$ at the $k^{\rm{th}}$ lens plane:  
\begin{equation}
\label{eq:shear_matrix_k_plane}
U_{ij}^{(k)}(\bm{\beta}) = \frac{\partial^2\psi^{(k)}(\bm{\beta})}{\partial \beta_i \partial \beta_j} =  \frac{\partial \alpha_{i}^{(k)}(\bm{\beta})}{\partial \beta_j} \, .
\end{equation}

We note that, since we are dealing with quantities on the curved sky, partial derivative operators are promoted to covariant derivatives \citep{Becker2013}. We have now introduced all the quantities needed for a numerical implementation of equations~(\ref{eq:grav_lens_eq}) and (\ref{eq:dist_matrix_def}). In particular, the integrals can be carried out as a discrete summation over $N$ spherical lens planes, which leads to the following expressions for $\bm \beta$ and $A$, respectively:
\begin{equation}
\label{eq:beta_discrete_sum}
\bm{\beta}^{(N)}(\bm{\theta}) = \bm{\theta} - \sum_{k=0}^{N-1} \frac{f^{(k,N)}_K }{f^{(N)}_K} \bm{\alpha}^{(k)} (\bm{\beta}^{(k)}) \, ,
\end{equation}
\begin{equation}
\label{eq:dist_matrix_discrete_sum}
A_{ij}^{(N)}(\bm{\theta}) = \delta_{ij} -  \sum_{k=0}^{N-1} \frac{f^{(k,N)}_K}{f^{(N)}_K} U_{il}^{(k)} A_{lj}^{(k)} \, ,
\end{equation}
where $f^{(k,N)}_K \equiv f_K(\chi^{(N)} - \chi^{(k)})$.

Using the above formulae with high-resolution maps would however require a large number of operations, and most importantly, the memory needs would be prohibitively large. \citet{Hilbert2009} reformulated these equations to allow the quantities to be calculated with an iterative procedure that requires only information about the previous two lens planes:
\begin{equation}
\label{eq:beta_discrete_sum_1}
\begin{aligned}
\bm{\beta}^{(k+1)} &= \left(1 - \frac{f^{(k)}_K}{f^{(k+1)}_K} \frac{f^{(k-1,k+1)}_K}{f^{(k-1,k)}_K}\right) \bm{\beta}^{(k-1)} 
\\ &+ \frac{f^{(k)}_K}{f^{k+1}_K} \frac{f^{(k-1,k+1)}_K}{f^{(k-1,k)}_K} \bm{\beta}^{(k)} - \frac{f^{(k,k+1)}_K}{f^{(k+1)}_K} \bm{\alpha}^{(k)}(\bm{\beta}^{(k)})
\, ,
\end{aligned}
\end{equation}
\begin{equation}
\label{eq:dist_matrix_discrete_sum_1}
\begin{aligned}
A_{ij}^{(k+1)} &= \left(1 - \frac{f^{(k)}_K}{f^{(k+1)}_K} \frac{f^{(k-1,k+1)}_K}{f^{(k-1,k)}_K}\right) A_{ij}^{(k-1)} 
\\ &+ \frac{f^{(k)}_K}{f^{k+1}_K} \frac{f^{(k-1,k+1)}_K}{f^{(k-1,k)}_K} A_{ij}^{(k)} - \frac{f^{(k,k+1)}_K}{f^{(k+1)}_K} U_{il}^{(k)} A_{lj}^{(k)}
\, .
\end{aligned}
\end{equation}
Here the initial conditions at the first lens plane are set to $\bm{\beta}^{(-1)} = \bm{\beta}^{(0)} = \bm{\theta}$ and $A_{ij}^{(-1)} = A_{ij}^{(0)} = \delta_{ij}$; i.e.~we perform backward-in-time ray-tracing from the observer to the source plane.

As noted by \citet{Becker2013}, when working with spherical lens planes, one has to parallel transport the distortion matrix (a tensor on the sphere) along the geodesic connecting the angular positions of the ray at consecutive planes, which takes into account the change in the local tensor basis.

\subsection{Spherical harmonics relations}
\label{subsec:sph}
Since we are working with full-sky maps, the optimal way to compute the derivatives in equations~(\ref{eq:deflection_k_plane}) and (\ref{eq:shear_matrix_k_plane}) is by performing them in the spin-weighted spherical harmonics domain \citep[see][for a standard reference]{Varshalovich}. In this section, we will discuss how we compute $\bm{\alpha}$ and $U$ at each lens plane by making use of the relations derived in \citet{Hu2000}. 

Once $\kappa$, a spin-0 (i.e.~scalar) quantity, is computed from equation~(\ref{eq:kappa_plane}) (in this section we omit the superscript $k$ for clarity), we can obtain its spherical harmonic coefficients $\kappa_{\ell m}$ using the HEALPix \texttt{map2alm} routine. The key relation is then:  
\begin{equation}
\label{eq:kappa_lm}
\kappa (\bm{\beta}) = \sum_{\ell=0}^{\infty} \sum_{m=-\ell}^{\ell} \kappa_{\ell m} Y_l^m (\bm{\beta}) \, ,
\end{equation}
where $Y_l^m$ are the spin-0 spherical harmonics. The Poisson equation on the surface of the sphere, i.e. equation~(\ref{eq:potential_k_plane}), takes the following form in the spherical harmonics domain: 
\begin{equation}
\label{eq:poisson_lm}
\kappa_{\ell m} = -\frac{\ell (\ell +1)}{2}  \psi_{\ell m} \, ,
\end{equation}
while the derivative in  equation~(\ref{eq:deflection_k_plane}) yields the spin-1 (i.e.~tangential vector) field:
\begin{equation}
\label{eq:alpha_lm}
\alpha_{lm} = - \sqrt{\ell (\ell +1)} \psi_{\ell m}  \, .
\end{equation}

We point out that one can skip the computation of the potential and obtain the deflection field directly from the convergence by combining equations~(\ref{eq:poisson_lm}) and~(\ref{eq:alpha_lm}):
\begin{equation}
\label{eq:kappa_alpha_lm}
\alpha_{lm} = \frac{2}{\sqrt{\ell (\ell +1)}} \kappa_{\ell m} \, .
\end{equation}

Regarding the shear matrix, we combine equations~(7) and~(8) of \citet{Castro2005}, yielding:
\begin{equation}
\label{eq:shear_matrix_kappa_gamma}
U_{ij}(\bm{\beta})  =   \kappa (\bm{\beta}) \delta_{ij} + [\gamma_1 (\bm{\beta}) \sigma_3 + \gamma_2 (\bm{\beta}) \sigma_1]_{ij} \, ,
\end{equation}
where $\sigma_1$ and $\sigma_3$ are the Pauli matrices. Therefore, the only additional quantity we need to obtain for the shear matrix is the shear, a spin-2 quantity which can be computed from $\kappa$ as follows:
\begin{equation}
\label{eq:gamma_lm}
\gamma_{lm} = - \sqrt{(\ell+2) (\ell-1) / (\ell (\ell+1))} \kappa_{\ell m} \, ,
\end{equation}
under the assumption that the B-modes of the shear (which have comparable power to the rotation) can be neglected \citep[see e.g.][]{Hadzhiyska2023c}.

\subsection{Interpolation on HEALPix maps}
\label{subsec:interpolation}

As one can see from equations~(\ref{eq:beta_discrete_sum_1}) and (\ref{eq:dist_matrix_discrete_sum_1}), $\bm{\alpha}^{(k)}$ and $U^{(k)}$ have to be evaluated at the angular position $\bm{\beta}^{(k)}$ for each ray, which in general will not coincide (apart from the first iteration) with the center of a HEALPix pixel. A classic approach to this problem consists of transforming the quantities back to real space with the HEALPix \texttt{alm2map\_spin} routine, and then performing interpolation in the resulting maps, with the simplest choices being Nearest Grid Point \citep[NGP, e.g.][]{Fabbian2018} and bilinear \citep[e.g.][]{Broxterman2024} interpolation. The first is faster but less precise, while the second is in principle more precise, but at the cost of being slower and introducing significant smoothing. We discuss the impact of the additional effective smoothing from bilinear interpolation in Appendix~\ref{appendix:interpolation}. 

A novel, more accurate and faster approach, based on the nonuniform fast Fourier transform (NUFFT, cf.\ \citealt{Fessler2003, Barnett2019}) algorithm, has been efficiently implemented in the context of gravitational lensing by \citet{Reinecke2023}. The idea is to accurately synthesize the fields at the desired angular positions by first computing a map on an equiangular grid from the spherical harmonic coefficients using a conventional spherical harmonic transform algorithm, then extending this map to $2\pi$ in the $\theta$ direction (which results in a map that is periodic along both coordinate axes), and finally interpolating the result to the desired locations using non-uniform FFTs. 

In this work we will compare NGP, bilinear and NUFFT interpolated ray-tracing simulations to the Born approximation. This will allow us to evaluate not only the impact of ray-tracing as opposed to the Born approximation, but also to capture and study the differences between the three different interpolation schemes employed.

\begin{figure*}
    \centering
    \includegraphics[width=0.96\textwidth]{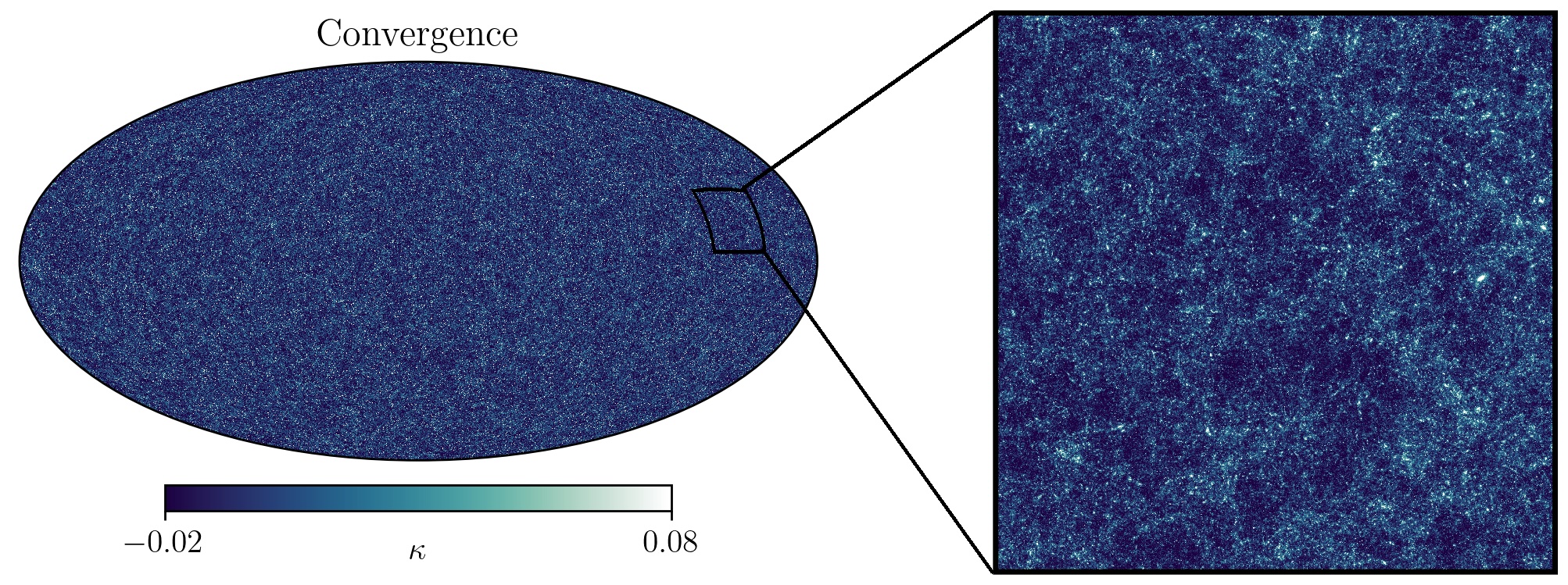}
    \includegraphics[width=0.96\textwidth]{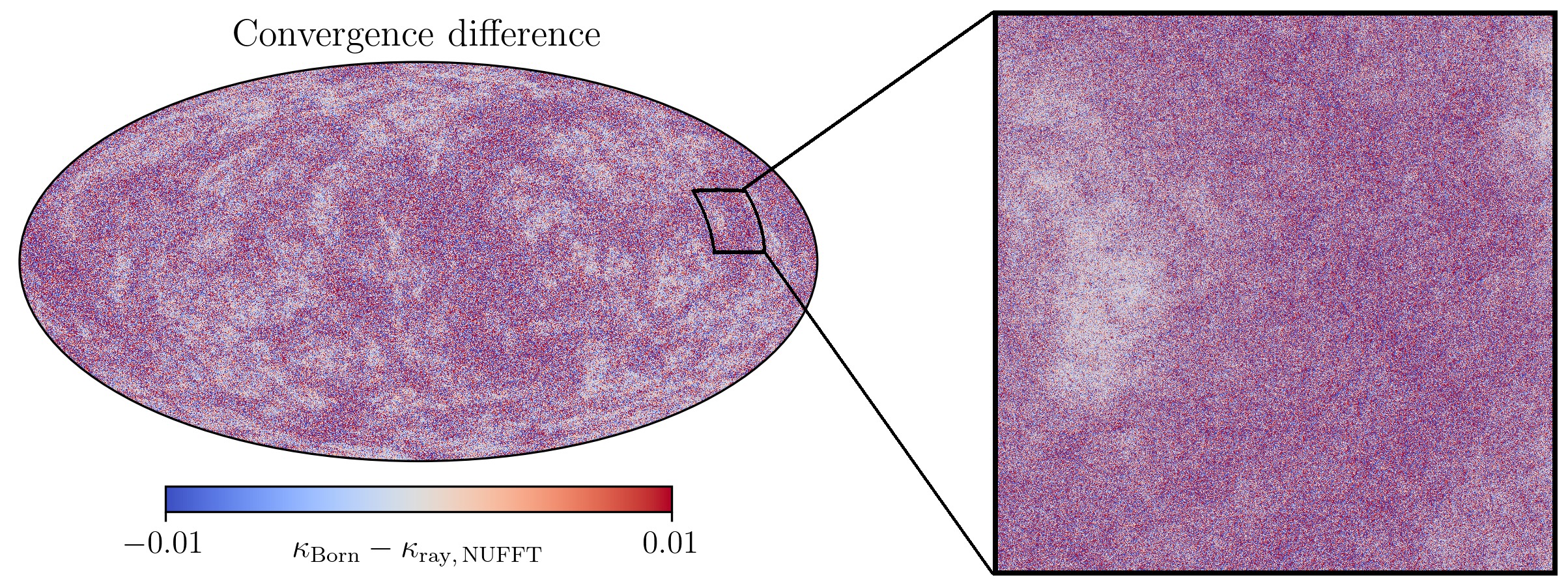}
    \includegraphics[width=0.96\textwidth]{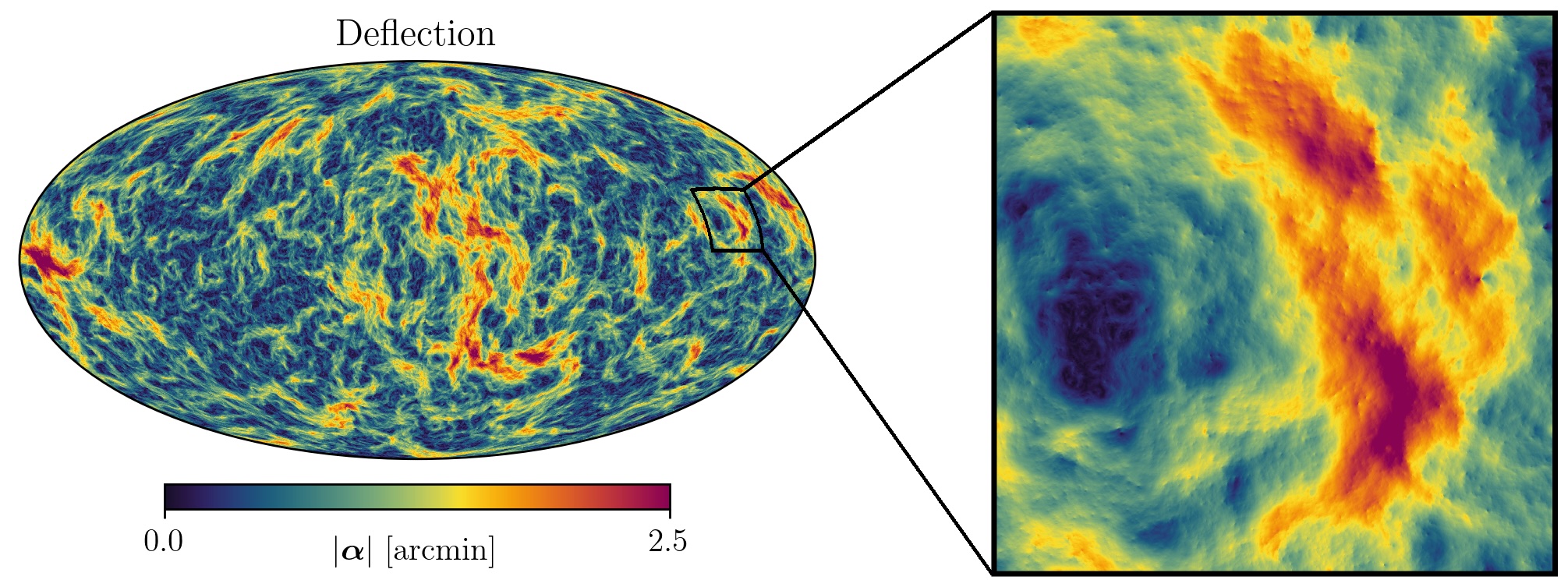}
    \caption{{\it Top row:} convergence field computed with our ray-tracing code using NUFFT interpolation. {\it Middle row}: difference of the convergence field when computed with ray-tracing instead of the Born approximation. {\it Bottom row:} amplitude of the deflection field $\bm{\alpha}$, computed as the difference between the observed angular position $\bm{\theta}$ and the original angular position $\bm{\beta}$. In each case, the panels on the right hand side show enlargements with a size of 10 degrees on a side. They may also be compared to the corresponding regions in the fields shown in Figure~\ref{fig:maps_with_zoom2}.}\label{fig:maps_with_zoom1}
\end{figure*}

\begin{figure*}
    \centering
    \includegraphics[width=0.96\textwidth]{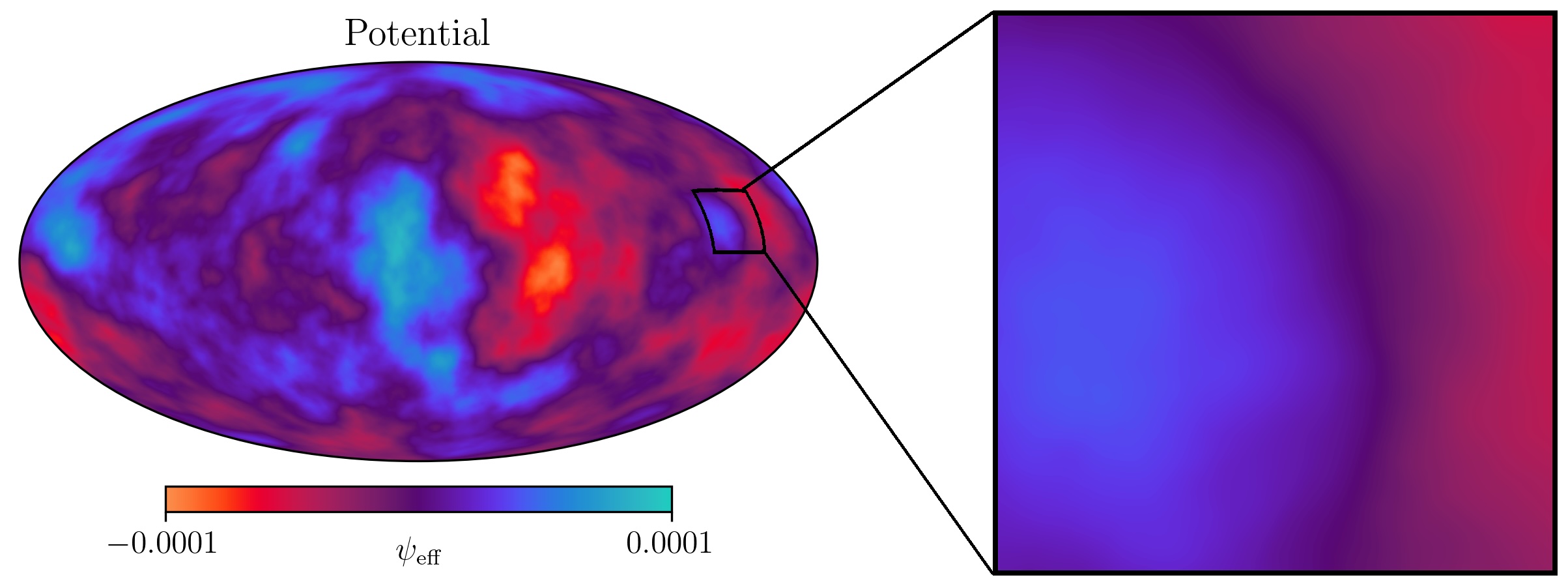}
    \includegraphics[width=0.96\textwidth]{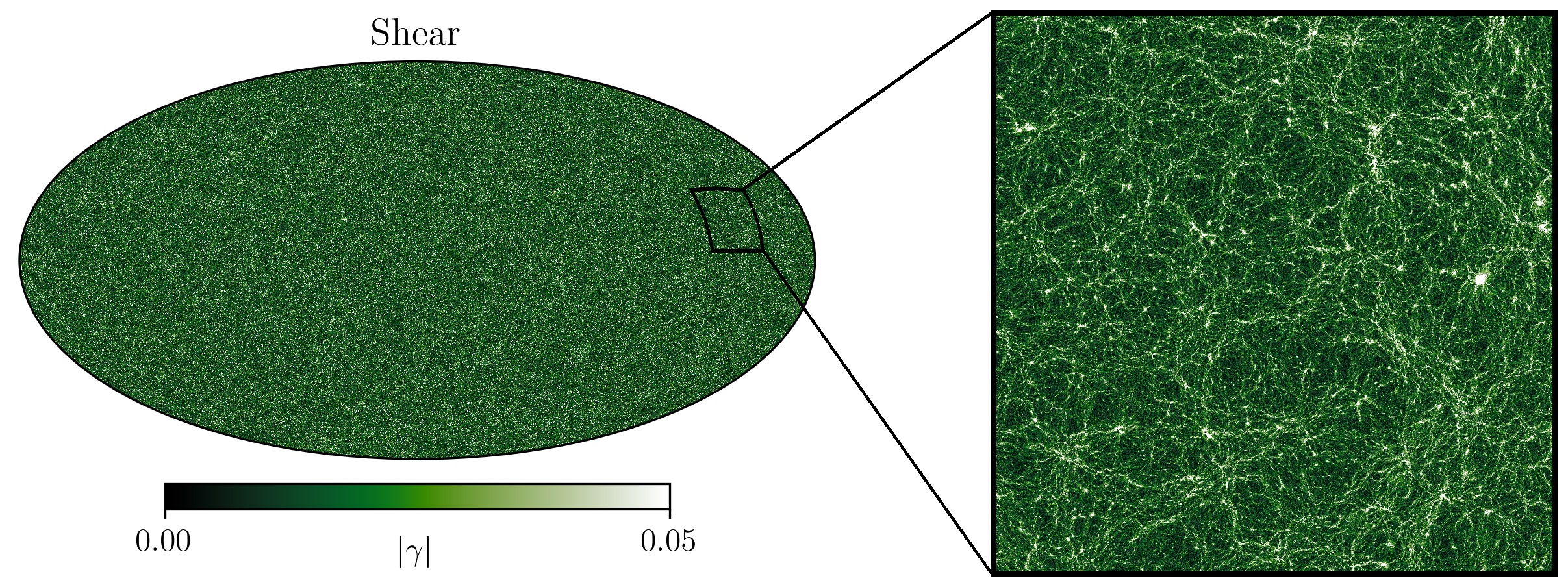}
    \includegraphics[width=0.96\textwidth]{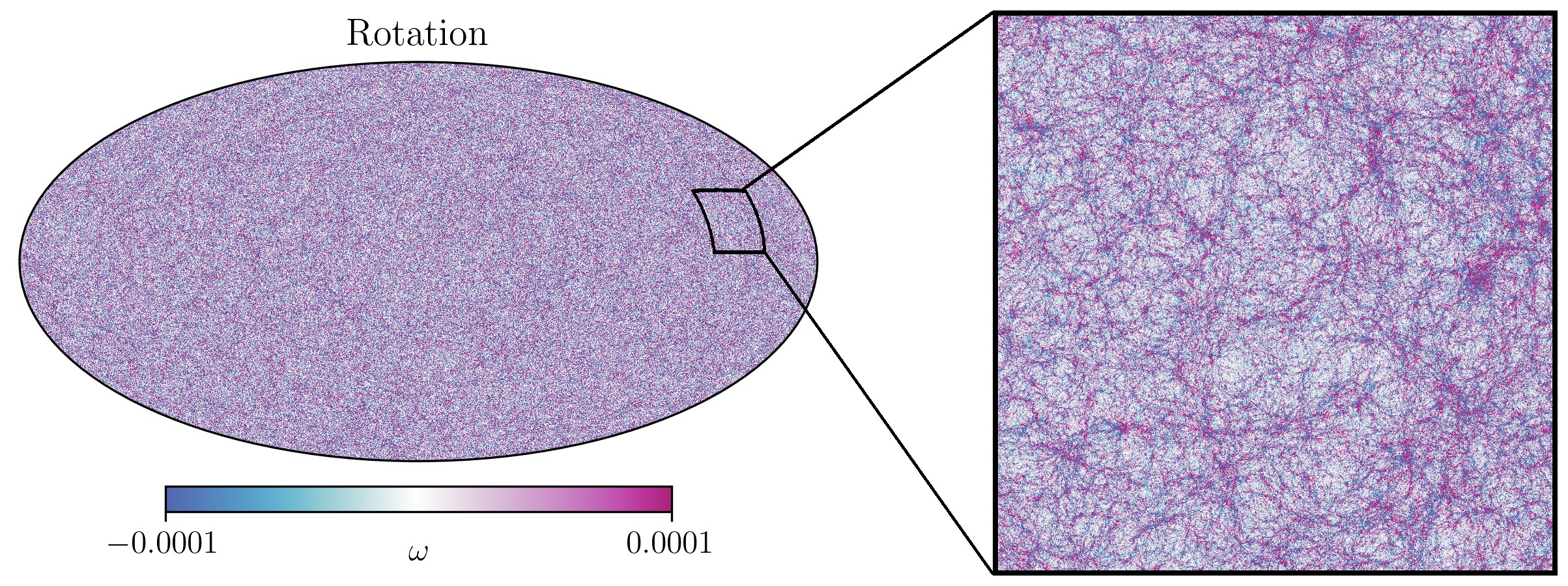}
    \caption{{\it Top row:} effective lensing potential, computed by plugging the convergence field into equation~(\ref{eq:poisson_lm}).  {\it Middle row}: shear field amplitude $|\gamma| = \sqrt{\gamma_1^2 + \gamma_2^2}$. {\it Bottom row:} rotation field. All quantities were computed with our ray-tracing code using NUFFT interpolation. In each case, the panels on the right hand side show enlargements with a size of 10 degrees on a side. They may also be compared to the corresponding regions in the fields shown in Figure~\ref{fig:maps_with_zoom1}. }\label{fig:maps_with_zoom2}
\end{figure*}

\subsection{Computation of the WL statistics}
\label{subsec:observables}

The power spectra are computed with the HEALPix routine \texttt{anafast},  and binned into 80 equally spaced logarithmic bins in the range $\ell \in [1,2.5 \times 10^4]$. Before computing all the other statistics, every map is smoothed with the HEALPix \texttt{smoothing} routine with a Gaussian symmetric beam characterized by a standard deviation of $1 \, \mathrm{arcmin}$, consistent with \citet{Ferlito2023} and with other similar studies. For the convergence PDF, we bin all pixels in 100 linearly spaced bins in the range $\kappa \in [-0.15, 0.25]$. Peaks and minima are computed as the pixels that are, respectively, greater or smaller than their 8 neighbours\footnote{In the HEALPix tessellation, every pixel has 8 neighbors, except for a small minority of pixels, for which it can be 7 or 6.}, which are retrieved using the HEALPix \texttt{get\_all\_neighbours} routine. For the peaks, we set 25 linearly spaced bins in the range $\kappa \in [-0.04, 0.2]$, while for the minima 30 linearly spaced bins in the range $\kappa \in [-0.045, 0.045]$ are used. Void statistics were computed following \cite{Davies2018}. In particular, for the void abundance, we set 25 linearly spaced bins in the range $R_v \in [0, 0.2] \, \mathrm{deg}$ where $R_v$ is the void radius, and for the stacked void profiles, we use 20 linearly spaced bins in the range $r/R_v \in [0, 2]$, where $r/R_v$ is the distance from the void center ($r$) in units of void radius. Finally, Minkowski functionals (MF) were computed with the publicly available Python package {\small PYNKOWSKI} \citep{Carones2024}, using 130 linearly spaced bins in the range $\kappa \in [-0.03, 0.11]$. For all of the above statistics, the value at each bin is computed as the average of the A- and B-realizations of our MTNG simulations. In the case of the void profiles and the third MF, in the lower sub-panel of the respective plots, we decided to show the difference with respect to the Born approximation. The reason being that since these two statistics cross the zero value, the ratio would diverge to infinity, making the plot somewhat more difficult to interpret. For all the other statistics, the lower sub-panel shows the ratio with respect to the Born approximation.


\section{Results}
\label{sec:results}

\begin{figure}
    \centering
    \includegraphics[width=0.48\textwidth]{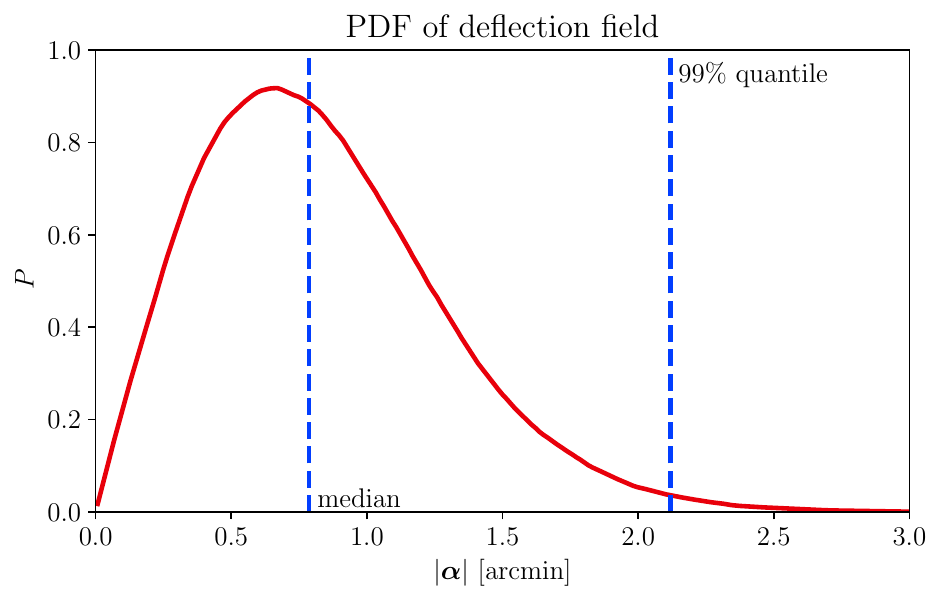}
    \caption{PDF of the deflection field  $\bm{\alpha}$, computed with our ray-tracing code using NUFFT interpolation as the difference between the observed angular position $\bm{\theta}$ and the original angular position $\bm{\beta}$. The dashed lines indicate the median and the 99\% percentile.}\label{fig:alpha_pdf}
\end{figure}

\begin{figure*}
    \centering
    \includegraphics[width=0.48\textwidth]{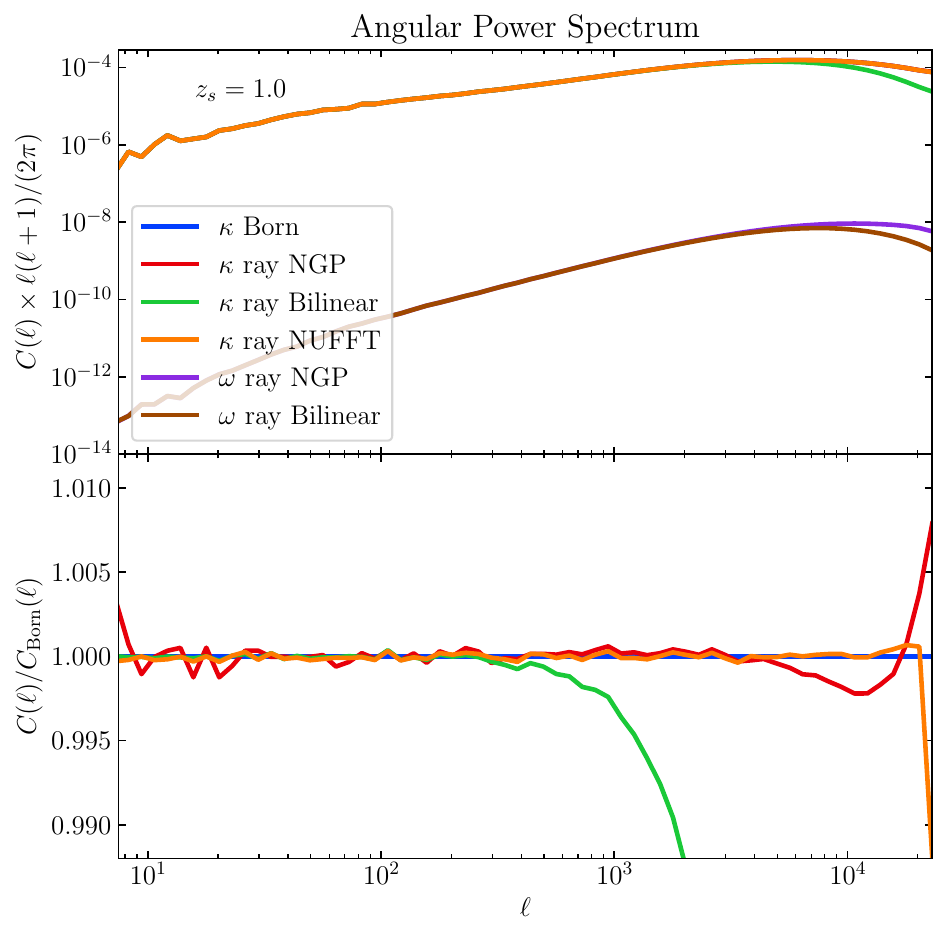}
    \includegraphics[width=0.48\textwidth]{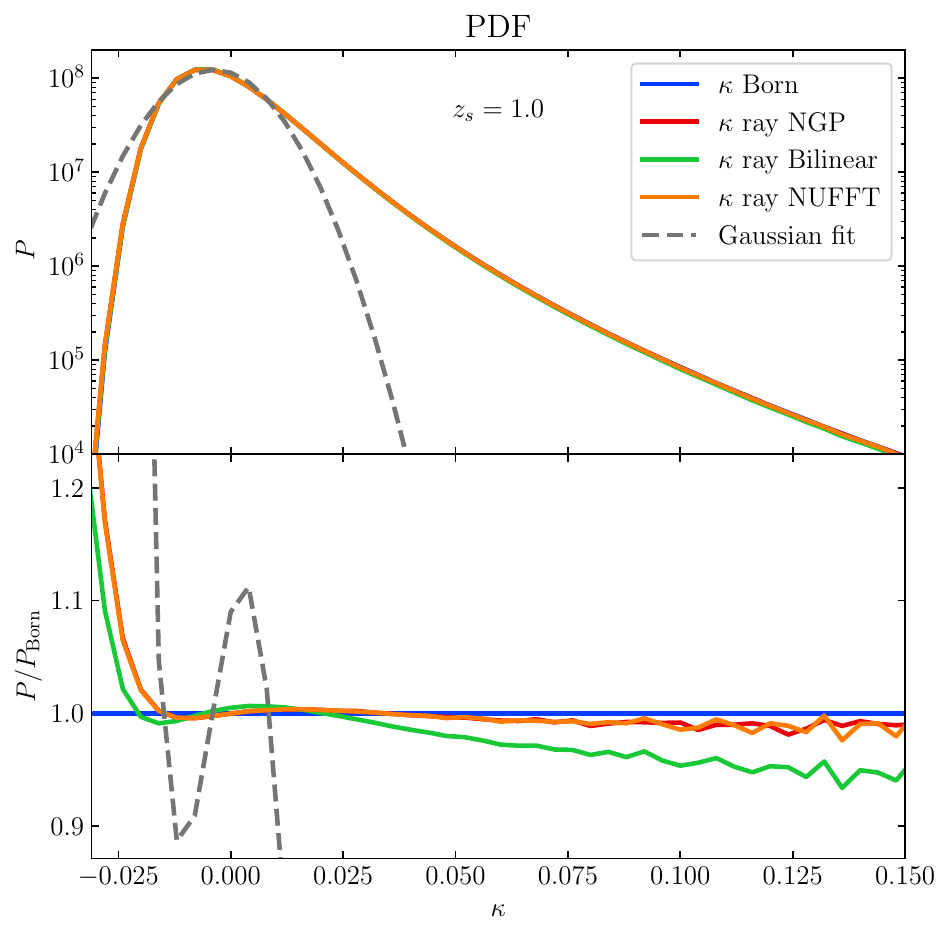}
    \includegraphics[width=0.48\textwidth]{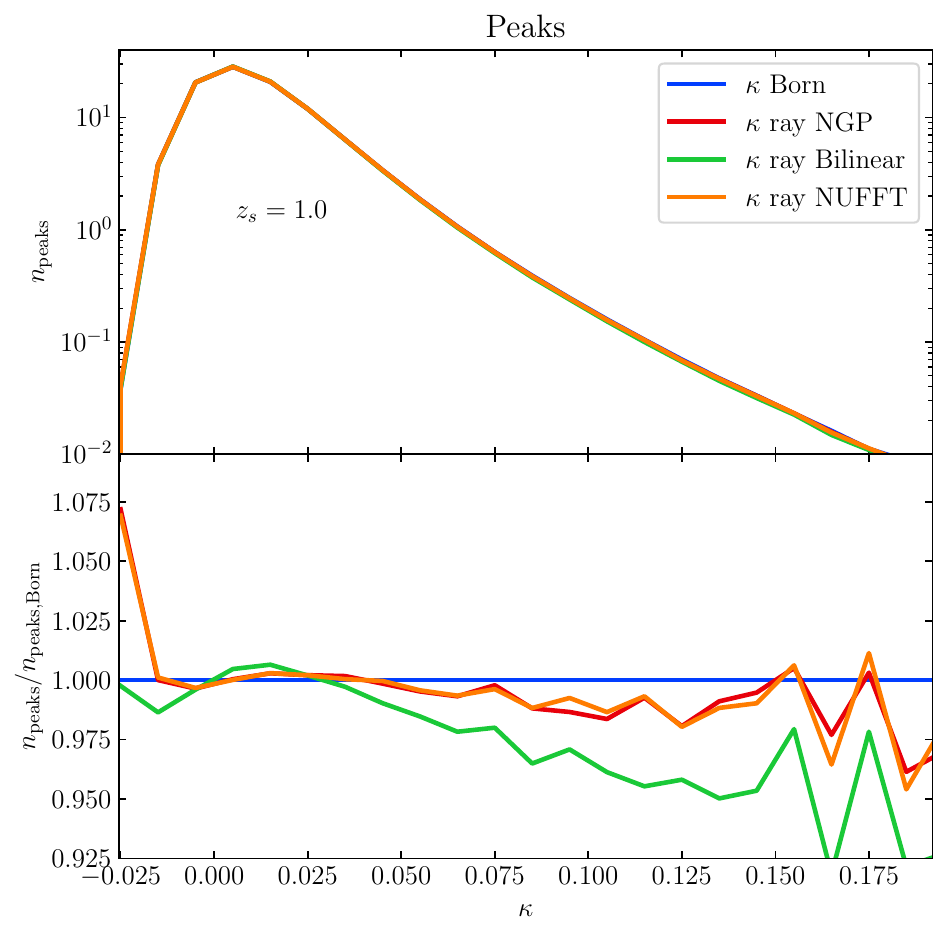}
    \includegraphics[width=0.48\textwidth]{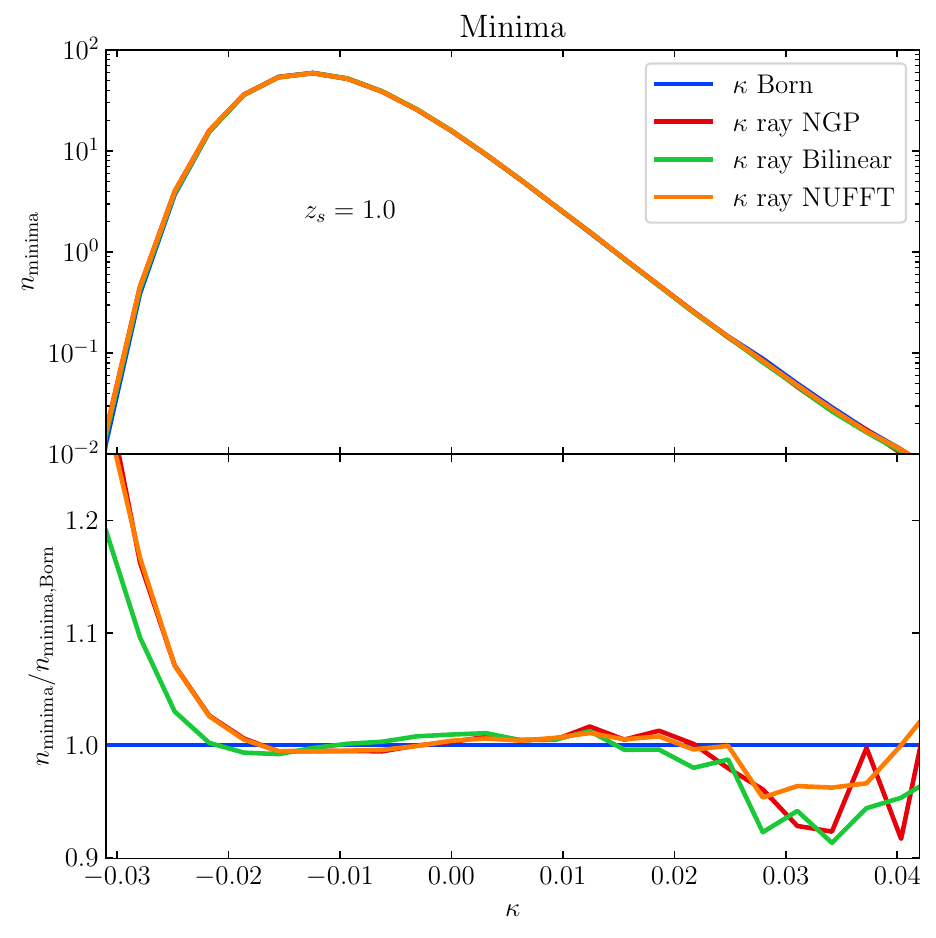}
    \caption{{\it Top left:} angular power spectrum; {\it top right:} PDF; {\it bottom left:} WL peak counts; {\it bottom right:} minimum counts. All the observables are computed adopting a fiducial source redshift of $z_{\rm s} = 1.0$. The statistics are extracted from WL convergence maps computed in the Born approximation (blue), as well as with our ray-tracing code, by using NGP (red), bilinear (green) and NUFFT (orange) interpolation. The lower sub-panel of each plot shows the ratio of these three ray-tracing results with respect to the Born approximation. In the case of the angular power spectrum, we also show the WL rotation computed with the ray-tracing code using NGP (purple) and bilinear (brown) interpolation. In the case of the PDF, we also include a Gaussian fit to the map obtained with the Born approximation (dashed grey). We observe that the power spectrum is not significantly affected by the Born approximation, while the convergence PDF is slightly Gaussianised by ray-tracing, and peaks and minima counts are impacted correspondingly. We also notice that, for all the observables, bilinear interpolation introduces a smoothing that significantly distorts the effects of ray-tracing itself.
    }\label{fig:born_vs_ray_pk_pdf_peaks_minima}
\end{figure*}

\begin{figure*}
    \centering
    \includegraphics[width=0.468\textwidth]{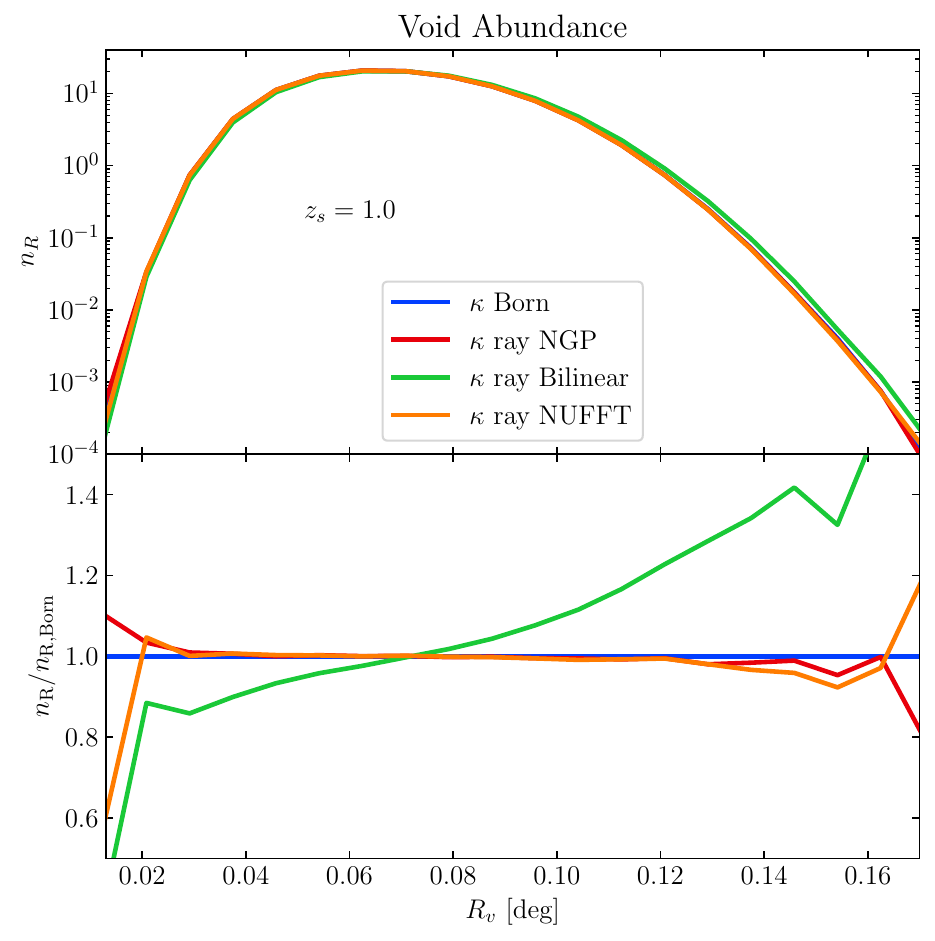}
    \includegraphics[width=0.492\textwidth]{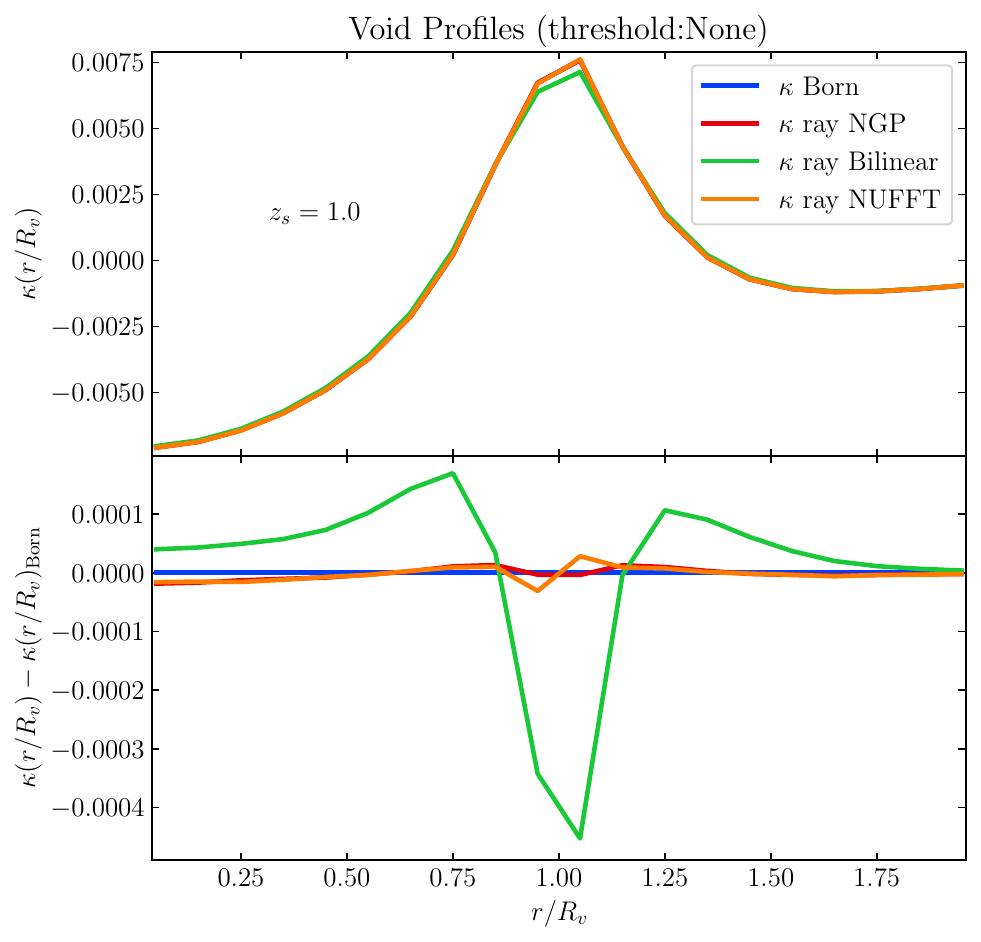}

    \caption{{\it Left:} void abundance distribution; {\it right left:} stacked void profiles. The statistics are extracted from WL convergence maps computed in the Born approximation (blue), as well as with our ray-tracing code, by using NGP (red), bilinear (green) and NUFFT (orange) interpolation. The maps have a source redshift of $z_{\rm s} = 1.0$. The lower sub-panel of each plot shows the ratio (or the difference, in the case of the profiles) of these three ray-tracing results with respect to the Born approximation. Neither statistic is significantly affected by the Born approximation. We also notice that bilinear interpolation introduces a smoothing that moves the void abundance distribution toward larger radii and flattens the void profiles.}\label{fig:born_vs_ray_voids}
\end{figure*}

\begin{figure}
    \centering
    \includegraphics[width=0.48\textwidth]{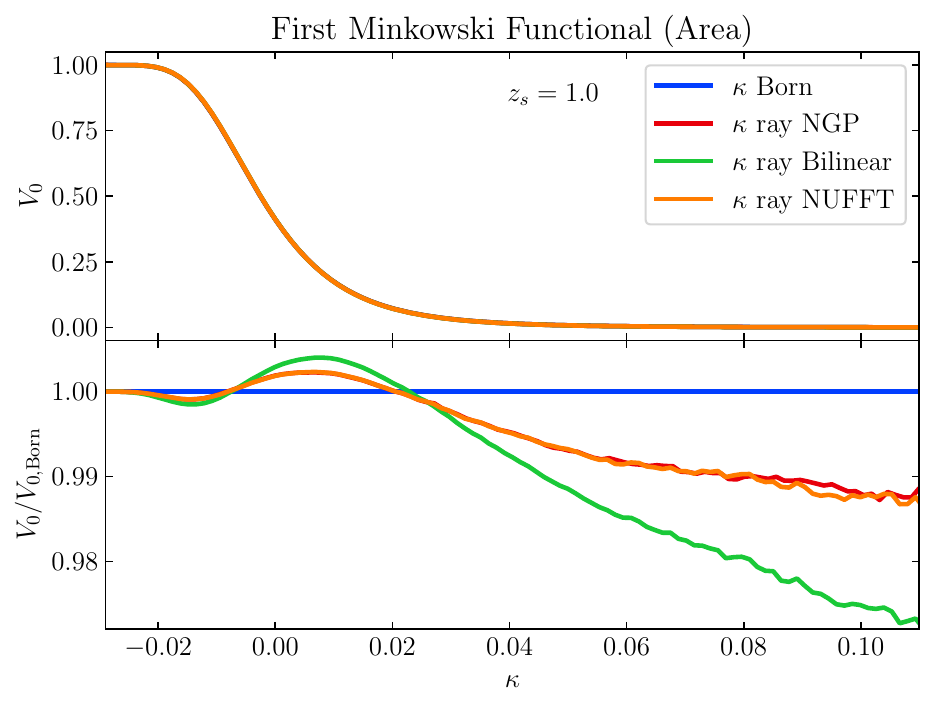}
    \includegraphics[width=0.48\textwidth]{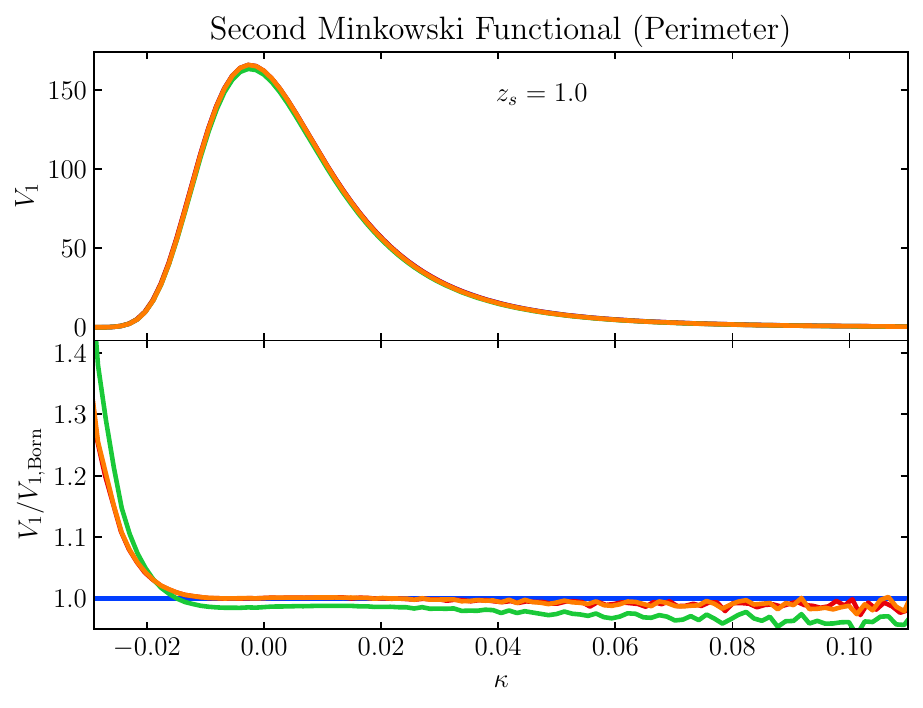}
    \includegraphics[width=0.48\textwidth]{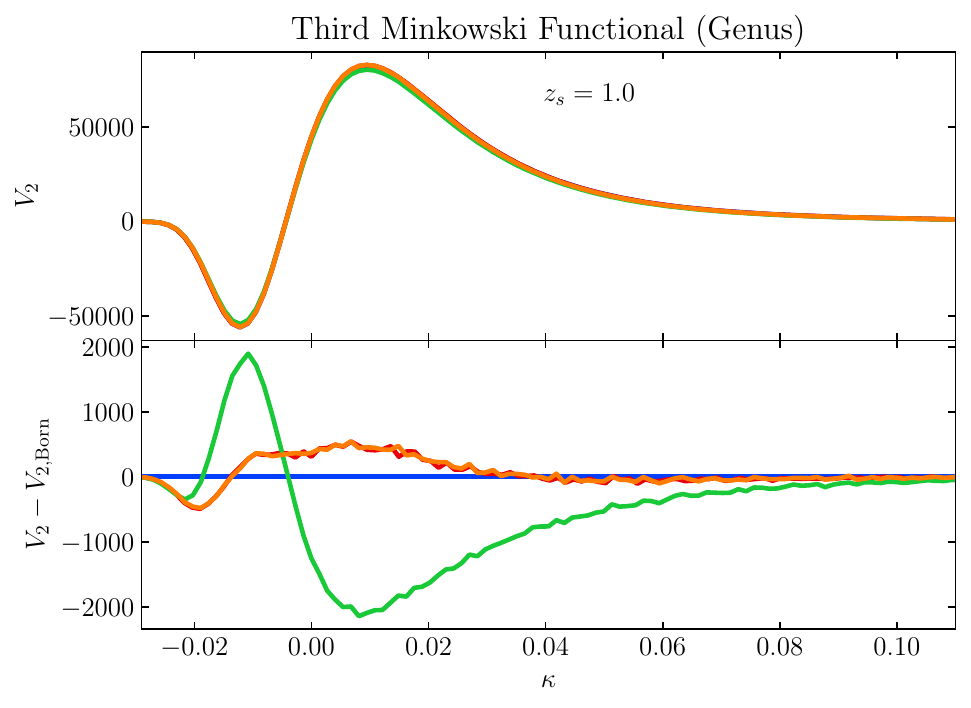}
    \caption{{\it Top:} first Minkowski functional (MF, area); {\it middle:} second MF (perimeter); {\it bottom:} third MF (genus). The statistics are extracted from WL convergence maps computed in the Born approximation (blue), as well as with our ray-tracing code, by using NGP (red), bilinear (green) and NUFFT (orange) interpolation. The maps have a source redshift of $z_{\rm s} = 1.0$. The lower sub-panel of each plot shows the ratio (or the difference, in the case of the genus) of these three ray-tracing results with respect to the Born approximation. We see that the three MFs are affected to different degrees and that the bilinear interpolation introduces a smoothing that significantly distorts the effects of ray-tracing itself.}\label{fig:born_vs_ray_mf}
\end{figure}

We begin the presentation of our results with an overview of some key quantities produced by our simulations, we then move our focus to the impact of the Born approximation on the following WL statistics: power spectrum, PDF, Peaks, Minima, Void statistics, and Minkowski functionals. For definiteness, all the WL statistics in the present work are computed assuming a source redshift of $z_{\rm s} = 1.0$.

\subsection{Overview of WL quantities}
\label{subsec:WL_quantities}

Figures~\ref{fig:maps_with_zoom1} and~\ref{fig:maps_with_zoom2} show maps of some essential WL quantities, as computed with our ray-tracing code using the NUFFT interpolation. We start by looking at the amplitude of the deflection field (bottom panel of Figure~\ref{fig:maps_with_zoom1}), which is defined as the difference between the observed angular position $\bm{\theta}$ and the original angular position $\bm{\beta}$. We see that the characteristic angular scale of fluctuations is larger than that of the convergence field (top panel of Figure~\ref{fig:maps_with_zoom1}): this is expected as the convergence is the derivative with respect to the angular position of the deflection. Features that cause the strongest deflection, i.e.~with an intensity $\gtrsim 2$ arcmin (colored in red), reach an angular size of $\approx 5-10$ degrees. This reflects the fact that for cosmological WL, the dominant component of the deflection field is generated from the large scale structure of the Universe, rather than single objects. To better quantify the magnitude of the deflection field, we can look at its PDF in Figure~\ref{fig:alpha_pdf}. We notice that the median deflection that a ray experiences along its path from the source to the observer is $\approx 0.79$ arcmin, which is almost twice the angular size of a pixel in our setting. Interestingly, only $1 \%$ of the rays experience a deflection greater than $\approx 2.1$ arcmin.

By comparing the full sky map of $|\bm{\alpha}|$ with the one of the convergence difference (center panel of Figure~\ref{fig:maps_with_zoom1}), we can perform a qualitative consistency check by noticing that the regions where the rays experience the least deflection (colored from dark blue to black) correspond to the regions where the difference between the Born approximation and ray-tracing is the closest to zero (colored from light grey to white). This becomes even more evident by looking at the zoomed-in region, where a low-deflection region is seen in the left half of the panel. 

In the top panel of Figure~\ref{fig:maps_with_zoom2}, we show the ``effective'' lensing potential $\psi_{\rm{eff}}$, computed by plugging the convergence field (the final product of our code) into equation~(\ref{eq:poisson_lm}). Consistent with theoretical expectations, a flatter potential will correspond to a smaller deflection (e.g. left half of the zoomed panel). Conversely, where the potential has strong variation, the deflection will be greater (e.g. right half of the zoomed panel). Analogously to the relation between convergence and deflection, we notice that the characteristic angular scales of fluctuations of the lensing potential are larger than those of the deflection, which is its derivative with respect to angular position.

In the center and bottom panels of Figure~\ref{fig:maps_with_zoom2}, we show the shear amplitude $|\gamma| = \sqrt{\gamma_1^2 + \gamma_2^2}$ and the rotation $\omega$, respectively. Both quantities are obtained from the distortion matrix $A$ computed with our ray-tracing code. By looking at the zoomed regions of convergence and shear, we can recognize the same underlying cosmic structure in both fields. But the two fields show strikingly different features: the convergence field is dominated by a huge number of single objects which will result in many peaks with different intensities; on the other hand, the shear intensity field shows a more interconnected structure, in which filaments and walls are clearly visible. Finally, we observe that the rotation field shows a very similar morphology to the one of the shear intensity field.

\subsection{Power Spectrum}
\label{subsec:powerspectrum}

In the top left panel of Figure~\ref{fig:born_vs_ray_pk_pdf_peaks_minima}, we show the angular power spectra of WL convergence and rotation. As expected from theoretical predictions, the power of the rotation field is $\sim$2-3 orders of magnitude smaller than the one of the convergence field. The most striking feature that can be observed for both quantities is the progressive power suppression at smaller angular scales. This suppression, observed when the ray-tracing scheme with bilinear interpolation is used, is around $\approx 1\%$ at $\ell \approx  1500$. As we discuss in more detail in Appendix~\ref{appendix:interpolation}, this effect is not directly connected to the ray-tracing, but rather arises when performing a series of bilinear interpolations on a HEALPix grid. 

Next, in the lower sub-panel, we show the ratio between the convergence spectra with ray-tracing and Born approximation. In the case of ray-tracing performed with NGP interpolation, we see that the fluctuations with respect to the Born approximation never exceed $ 0.2\%$, at least until $\ell \approx 10^4$  where we see a suppression of $0.25\%$, followed by a steep increase in power.  This effect, which occurs at approximately the pixel scale, could also be connected to the interpolation scheme. We have verified this by checking that this feature moves to lower multipoles when we decrease the angular resolution of the corresponding HEALPix maps. Finally, in the case of ray-tracing performed with NUFFT interpolation, we see that fluctuations never exceed $ 0.1\%$ at every scale, until the very last bin, at $\ell \approx 2.3 \times 10^4$, where a $\approx 1\%$ suppression is found.

We conclude that, modulo some $\lesssim 0.2\%$ fluctuations \citep[consistent with][]{Hilbert2020}, the effects on the power spectrum when performing ray-tracing instead of the Born approximation are dominated by the interpolation schemes, rather than from the improved modeling of the physical process. For this reason, NUFFT is found to be a preferrable scheme, as it introduces the least power distortions.

\subsection{Convergence PDF}
\label{subsec:pdf}

Having verified that ray-tracing has a negligible impact with respect  to the Born approximation at the level of the power spectrum, we next investigate whether this is also true for higher-order statistics. We start with the convergence one-point PDF, shown in the top right panel of  Figure~\ref{fig:born_vs_ray_pk_pdf_peaks_minima}. By looking at the lower sub-panel, which shows the ratio of the ray-tracing schemes to the Born approximation, we see that for NGP and NUFFT the central region is minimally distorted, with deviations never exceeding $\approx 1\%$ in the range $-0.015 < \kappa < 0.07$. Conversely, the outer regions of the distribution exhibit two opposite trends: for $\kappa \gtrsim 0.04$ there is a weak but progressive suppression that reaches $\approx 1.5\%$ at $\kappa \approx 0.09$; for $\kappa \lesssim -0.015$, we see a steep enhancement that reaches $\approx 10\%$ at $\kappa \approx -0.025$. In the case of ray-tracing with bilinear interpolation, there is a larger suppression of the high-$\kappa$ tail, which reaches $\approx 4\%$ at $\kappa \approx 0.09$, while the steep upturn in the low-$\kappa$ tail is shifted to smaller values. The discrepancy between this last ray-tracing method and the previous two can be explained by taking into account the smoothing introduced by the bilinear interpolation, which narrows the PDF. Our conclusion is that the Born approximation is likely to mildly overestimate the high-$\kappa$ tail and significantly underestimate the low-$\kappa$ part of the PDF. 

In contrast to the power spectrum, the features here are not dominated by the interpolation schemes. In particular, a smoothing of the PDF will lead to a suppression at negative $\kappa$, whereas we see the opposite in this case. Additionally, while for the power spectrum the three interpolation schemes have noticeably different effects, for the PDF the impact is similar in all cases, indicating that this is driven by a common underlying process.

An interpretation of this result can be given by referring to the leading-order post-Born corrections introduced in equation~(\ref{eq:kappa_quadratic}). Both the geodesic correction and the lens-lens coupling contribute to making the convergence distribution more Gaussian. This can be directly seen by considering the Gaussian fit (dashed grey line). The former does this by repeatedly displacing the ray positions along directions tangential to the line of sight. The latter by progressively processing the light signal through a system of lenses that are not correlated on sufficiently large scales.

We note that our results found in the case of NGP interpolation are qualitatively consistent with the numerical study of \citet{Fabbian2018} and the theoretical work of \citet{Barthelemy2020}. While they found somewhat stronger effects, this is to be expected as they studied CMB lensing.

\subsection{Peaks and minima}
\label{subsec:peaks_minima}

In the bottom left and bottom right panels of Figure~\ref{fig:born_vs_ray_pk_pdf_peaks_minima} we show our results regarding peaks and minima counts. In the case of ray-tracing with bilinear interpolation, we observe for both a uniform $\approx 2.5 \%$ suppression of the counts of peaks and minima at all values of $\kappa$. For a better comparison, this effect was removed in our figures by normalizing the count distributions to unity. 

Let us first discuss the peak counts. We see that the overall results are qualitatively similar to what we found for the PDF. In this case, the most relevant effect is the suppression of the high-$\kappa$ tail which, in the case of the ray-tracing with NGP and NUFFT interpolation, reaches $\approx 1.5\%$ at $\kappa \approx 0.15$. The additional suppression in the tails of the distribution in the case of ray-tracing with bilinear interpolation is explained by the smoothing that such a scheme introduces.

In the case of the minima counts, we observe that both tails of the distribution are distorted quite significantly. In particular, in the case of ray-tracing with NGP and NUFFT interpolation, the low-$\kappa$ tail experiences a steep enhancement amounting to $\approx 15\%$ at $\kappa \approx -0.025$, while the high-$\kappa$ tail is suppressed by $\approx 7\%$ at $\kappa \approx 0.03$. Similarly to the case of the PDF, ray-tracing with bilinear interpolation introduces a delay in the enhancement of the low-$\kappa$ tail, which can again be explained in terms of the smoothing introduced by the interpolation. 

Overall, it is interesting that the effects of Gaussianisation induced by ray-tracing are stronger for minima. This appears clearer by noting that the red and orange curves (NGP and NUFFT respectively) are the ones deviating more from the Born approximation, while, in the case of the peaks, it is the green curve (bilinear) that has the stronger deviation, indicating that the smoothing from bilinear interpolation is dominating over the post-Born effects.

\subsection{Void statistics}
\label{subsec:voids}

We continue our investigation by considering WL tunnel voids, a WL higher-order statistic that is sensitive to extended underdense regions and has been shown to have promising constraining power on cosmological parameters \citep{Davies2021}. The voids are defined as large underdense regions in the convergence field. The void-finding method used here is the tunnel algorithm, which identifies the largest circles that are empty of suitably defined tracers. In the present case, the tracers are the WL peaks of the convergence field. The corresponding void abundance and profiles are shown in the left and right panels of Figure~\ref{fig:born_vs_ray_voids}, respectively.

From the void abundance, we find that ray-tracing with bilinear interpolation distorts the distribution by shifting it towards larger void radii $R_v$, which is consistent with the induced additional smoothing. In particular, we observe a $\approx 20\%$ suppression in the low radii tail at $R_v \approx 0.02$ deg, and a $\approx 40\%$ enhancement in the high radii tail at $R_v \approx 0.14$ deg. In the case of ray-tracing with NGP and NUFFT interpolation, we only find a much smaller effect, namely a suppression at $R_v \approx 0.15$ deg amounting to around $\approx 10\%$, and deviations at the most extreme bins of the distribution, at $R_v < 0.02$ deg and $R_v < 0.16$ deg.

Moving to the stacked radial void profiles, we find that also in this case ray-tracing with bilinear interpolation has the strongest impact: it enhances the inner and outer regions while suppressing intermediate radii by $\approx 5\%$, at the peak of the convergence. This flattening of the profile can be explained by the effective smoothing introduced by bilinear interpolation. The effect of ray-tracing with NGP and NUFFT interpolation is significantly weaker, with the most noticeable feature being a $\approx 0.5\%$ suppression of the convergence in the innermost regions of the voids. This effect can be connected to the enhancement we observe in the low tail of the PDF of the minima distribution. In the case of NUFFT interpolation, we observe additional $\approx 1\%$ deviations at $r/R_v \approx 1$.

In general, we find that, modulo the artificial smoothing effects introduced by the bilinear interpolation, the post-Born corrections do not have a significant impact on void statistics. We tested the above statistics for a range of peak catalogue thresholds \citep{Davies2018} and found qualitatively similar results.

\subsection{Minkowski Functionals}
\label{subsec:minkowski}

Finally, we investigate the impact of the Born approximation on the Minkowski functionals. These are in general a set of $N+1$ morphological descriptors invariant under rotations and translations, that characterize a field in an $N$-dimensional space. The MFs are defined on an excursion set $\Sigma (\nu) = {\kappa > \nu \sigma_0}$; i.e.~for the set of pixels whose values exceed a certain threshold $\nu \sigma_0$. Here $\sigma_0$ is conventionally chosen as the standard deviation of the field. 

In our case we are dealing with a 2D field on the surface of the sphere, therefore we have the following three MFs: $V_0$ is simply the total area of the excursion set; $V_1$ is one-fourth of the total perimeter of the excursion set; and finally $V_2$, called ``genus'', is associated with the number of connected regions minus the number of topological holes of the excursion set. We refer the reader to equations~(2.5), (2.6) and (2.7) of \citet{Marques2024} for the mathematical expressions of $V_0$, $V_1$ and $V_2$. Since MFs encode information from all the moments of the distributions of a field, they are highly sensitive to non-Gaussianities and have thus been proposed as a powerful cosmological statistic in a number of different studies \citep[see e.g.][]{Springel1998, Schmalzing1998, Hikage2008, Ducout2013}.

We show our results for the $V_0$, $V_1$ and $V_2$ functionals in the top, middle and bottom panels of Figure~\ref{fig:born_vs_ray_mf}, respectively. In general, also in this case we note that any discrepancy between bilinear interpolation (green curve) with respect to the other two methods, NGP and NUFFT (red and orange curves), helps us to disentangle the impact of an additional smoothing introduced by bilinear interpolation, from the impact of post-Born corrections themselves. 

Let us recall that, by definition, $V_0$ is the cumulative PDF. Indeed, the impact of the three ray-tracing schemes can be directly related to what we observed for the PDF earlier (top right panel of Figure~\ref{fig:born_vs_ray_pk_pdf_peaks_minima}). All of the ray-tracing schemes show qualitatively the same trend, with an initial tiny suppression, followed by a small enhancement, and then by a stronger suppression. At $\kappa \approx 0.1$ this reaches $\approx 2.5\%$ for the bilinear interpolation, and $\approx 1.2\%$ for the NGP and NUFFT interpolation. 

In the case of $V_1$, at $\kappa \approx -0.028$, we see an enhancement of $\approx 35\%$ for ray-tracing with bilinear interpolation and of $\approx 25\%$ in the case of NGP and NUFFT. With increasing values of $\kappa$ there is a progressive suppression that, at $\kappa \approx 0.1$, reaches $\approx 4\%$ for bilinear interpolation and $\approx 1\%$ for NGP and NUFFT.

In the case of $V_2$, the genus statistic, we notice that ray-tracing with bilinear interpolation results in notably different effects: we see an enhancement of $\approx 3.5\%$ at $\kappa \approx -0.01$, followed by a suppression of $\approx 2.5\%$ at $\kappa \approx 0.01$ that then progressively diminishes, and eventually fades completely. For NGP and NUFFT interpolation, the effects are smaller. These are a suppression of $\approx 3\%$ at $\kappa \approx -0.02$, followed by an enhancement of $0.5 \%$ in the range $-0.015 \lesssim \kappa \gtrsim 0.04$.

Also in this case, the effects of smoothing by bilinear interpolation dominate the post-Born corrections. In particular, as expected, smoothing will shift the distribution to lower thresholds for the perimeter, while for the genus, it will dampen the amplitude.


\section{Conclusions and outlook}
\label{sec:conclusions}

In this paper, we present our methodology for computing full-sky ray-traced weak lensing maps, starting from the mass-shell outputs of {\small GADGET-4} and applying our code to a subset of the MillenniumTNG simulation suite. After having qualitatively inspected key WL quantities such as convergence, deflection, shear, and rotation, we test the impact of the Born approximation against three ray-tracing schemes that employ NGP, bilinear and NUFFT interpolation, respectively. These tests were performed on the power spectrum, as well as on a number of popular higher-order WL statistics.

We confirm, in line with theoretical predictions, that post-Born effects tend to Gaussianise the convergence PDF and consequently impact higher-order statistics as well. Regarding the use of different interpolation approaches in ray-tracing schemes, we interestingly find that although bilinear interpolation is in principle more accurate than NGP, the effective smoothing that this introduces at our grid resolution dominates the post-Born effects, even when using a HEALPix map with the maximum resolution of $N_{\rm side}=8192$. Additionally, we find that NUFFT interpolation, which is the most accurate method, agrees well with the NGP scheme in the present study. We can explain this by noting that a 1 arcmin smoothing, applied before the computation of all the higher-order statistics, tends to wash out information at the smallest angular scales, where the differences between these two methods are expected to become appreciable. We note that since the accuracy of NGP interpolation strongly depends on the HEALPix resolution, we expect the agreement with NUFFT to be degraded for lower values of $N_{\rm side}$.

We summarize the impact of using a ray-tracing scheme instead of the Born approximation for the different statistics as follows:
\begin{itemize}
    \item The angular power spectrum is not significantly affected by the Born approximation. Smoothing due the use of bilinear interpolation in a ray-tracing approach suppresses the power at progressively smaller scales.
    \item The convergence PDF is slightly Gaussianised by ray-tracing, which enhances the low-$\kappa$ tail and suppresses the high-$\kappa$ tail. We find deviations at $\kappa \approx -0.025$ of $\approx 10\%$ when NGP and NUFFT interpolation are used, and of $\approx 4\%$  and bilinear interpolation are used.
    \item Peaks counts are suppressed, at $\kappa \simeq 0.15$, by $\approx 1.5\%$ for ray-tracing with NGP and NUFFT interpolation and by $\approx 2.5\%$ for bilinear interpolation.
    \item Minima counts on the other hand are mainly enhanced, at $\kappa \approx -0.025$, by $\approx 15\%$ for NGP and NUFFT interpolation and by $\approx 7\%$  for bilinear interpolation.
    \item The void abundance and the void profiles are not significantly influenced by the Born approximation. However, smoothing due to bilinear interpolation distorts the void abundance with deviations up to $\approx 40\%$, and it flattens the void profiles with deviations up to $\approx 5\%$.
    \item The three 2D Minkowski functionals we investigated are affected to different degrees. Most noticeably, $V_1$ is enhanced at $\kappa \approx -0.028$ by $\approx 35\%$ for ray-tracing with bilinear and by $\approx 25\%$  for ray-tracing with NGP and NUFFT interpolation.
\end{itemize}

Overall, we find only very subtle consequences due to the use of the Born approximation. However, for higher-order statistics, they become sizable enough that the use of ray-tracing is necessary if exquisite precision is required. The need for interpolation arising in ray-tracing schemes typically introduces the technical problem of additional discreteness effects that can diminish or even defeat the accuracy improvements that ray-tracing in principle offers. Such problems can be overcome by adopting a novel interpolation scheme based on NUFFT. For currently achievable all-sky HEALPix resolutions, we find that low-order NGP interpolation agrees well with the NUFFT scheme, especially when a 1 arcmin smoothing is adopted. On the other end, the bilinear interpolation is found to be unreliable as it introduces sizable smoothing effects. Ultimately, in the limit of infinite resolution, we would expect the three interpolation schemes to agree. 

Note that the impact of post-Born corrections will increase with increasing source redshift. For WL surveys, source redshift distributions are smooth functions that vary between $z=0$ up to approximately $z \approx 3$, peaking at $z \approx 1$. In this work, our source redshift distribution corresponds to a $\delta$-function at $z=1$, which is roughly the median source redshift of typical Stage-IV WL surveys. Therefore, we expect the results presented here to be indicative of what can be expected for more realistic redshift source distributions.

We conclude that harvesting the accuracy benefits of ray-tracing is ultimately {\it required} in full-sky WL simulations in order to accurately model higher-order statistics to the percent level. 

The methods presented and tested in this work pave the way to several applications in the context of high-fidelity modeling in the era of precision cosmology. One possible development could be to compute, from the simulations used in this work, a detailed and realistic full-sky galaxy mock catalogue based on the latest version of the semi-analytic galaxy formation code {\small L-GALAXIES} \citep[][]{Barrera2023} which can be then combined with our ray-tracing code to obtain highly accurate predictions for the 3x2pt as well as higher order statistics. Another possibility consists of using the full-hydro MTNG run to extract the galaxy intrinsic alignment signal (a key contaminant of cosmological WL) from the lightcone \citep[in a similar way to][]{Delgado2023} and to compare/combine it with the shear signal computed with our ray-tracing code.

\section*{Acknowledgements}

FF would like to thank Soumya Shreeram, Ken Osato, William Coulton, Boryana Hadzhiyska, Javier Carrón Duque and Giulio Fabbian for useful discussions. For carrying out the MillenniumTNG simulations, the authors gratefully acknowledge the Gauss Centre for Supercomputing (GCS) for providing computing time on the GCS Supercomputer SuperMUC-NG at the Leibniz Supercomputing Centre (LRZ) in Garching, Germany, under project pn34mo. The work also used the DiRAC@Durham facility managed by the Institute for Computational Cosmology on behalf of the STFC DiRAC HPC Facility, with equipment funded by BEIS capital funding via STFC capital grants ST/K00042X/1, ST/P002293/1, ST/R002371/1 and ST/S002502/1, Durham University and STFC operations grant ST/R000832/1. 
VS and CH-A acknowledge support from the Excellence Cluster ORIGINS which is funded by the Deutsche Forschungsgemeinschaft (DFG, German Research Foundation) under Germany's Excellence Strategy -- EXC-2094 -- 390783311. VS and LH also acknowledge support by the Simons Collaboration on ``Learning the Universe''. SB is supported by the UK Research and Innovation (UKRI) Future Leaders Fellowship (grant number MR/V023381/1).

\section*{Data Availability}

The ray-tracing code {\small DORIAN} is publicly available on GitLab\footnote{\url{https://gitlab.mpcdf.mpg.de/fferlito/dorian}}. The simulations of the MillenniumTNG project are foreseen to be made fully publicly available in 2024 at the following address: \url{https://www.mtng-project.org}. The data underlying this article is available upon reasonable request to the corresponding author.

\bibliographystyle{mnras}
\bibliography{WL_raytracing}

\appendix

\begin{figure*}
    \centering
    \includegraphics[width=0.48\textwidth]{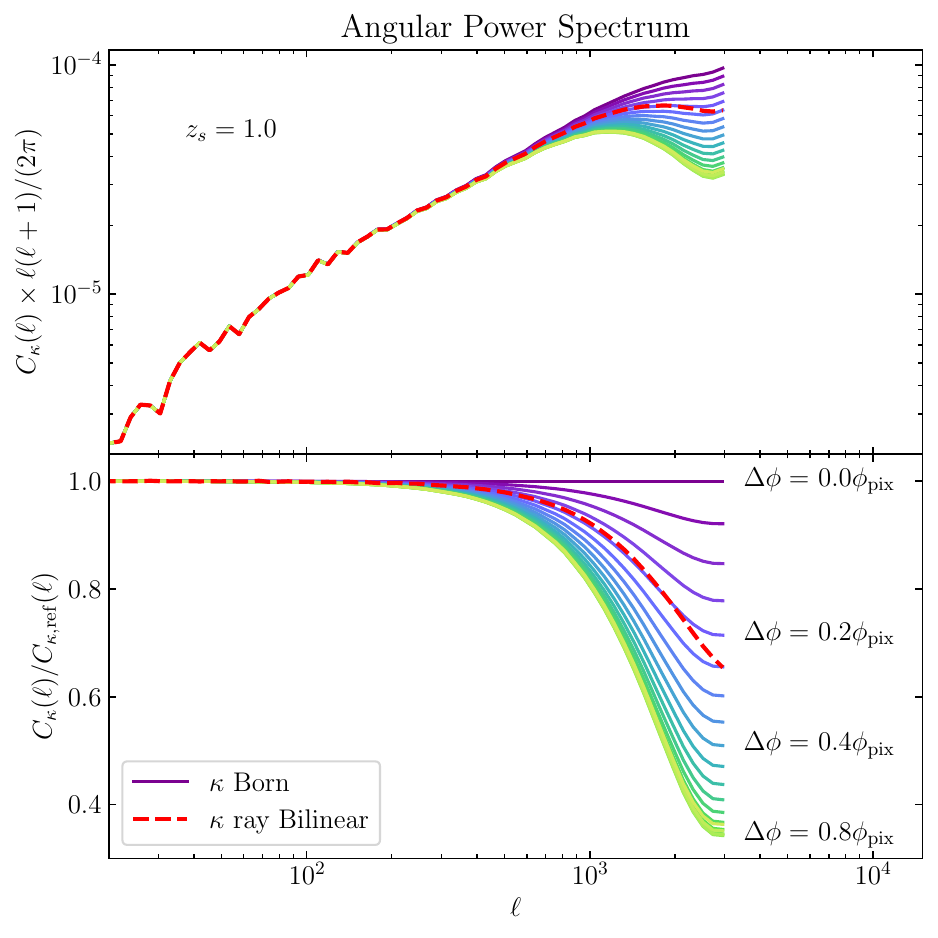}
    \includegraphics[width=0.48\textwidth]{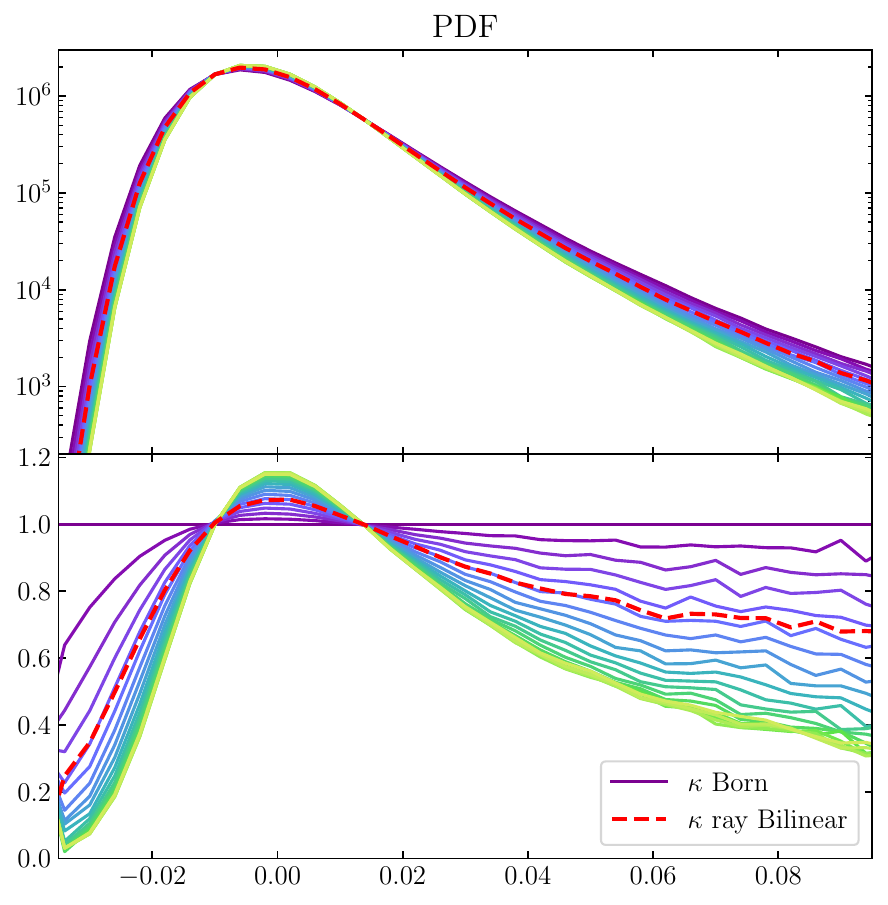}
    \caption{{\it Left panel:} angular power spectrum; {\it right panel}: PDF of the convergence field. These observables are computed for a fiducial source redshift of $z_{\rm s} = 1.0$. The solid lines indicate maps that were computed by interpolating from a convergence map in the Born approximation, but using a grid whose pixel centers were shifted longitudinally by an angle $\Delta \phi$, which was systematically varied from 0 to the pixel size $\phi_{\rm{pix}}$, as labeled. The red dashed line indicates convergence maps computed with ray-tracing and bilinear interpolation, for comparison.}\label{fig:pk_pdf_bilinear_interp}
\end{figure*}

\section{Impact of bilinear interpolation}
\label{appendix:interpolation}

In the present work, one of our ray-tracing setups features bilinear interpolation on HEALPix maps, which introduces a smoothing that progressively suppresses the power on small scales. This also narrows the PDF of the convergence, as well as the peaks and minima counts distributions.  

In the case of a field like the convergence, in which values can vary drastically from pixel to pixel, interpolating on points that are far from pixel centers introduces a significantly stronger smoothing with respect to points close to pixel centers.  To better quantify this effect we performed the following numerical test. We start with a convergence map, based on a HEALPix grid with $N_{\rm side}=1024$. We then computed a new map by performing a bilinear interpolation on the original map where we effectively rotated the underlying HEALPix grid along the equator by $\Delta \phi$, a fraction of the pixel angular size $\phi_{\rm{pix}}$. By repeating the above operation with increasing values of $\Delta \phi$, we interpolated on angular positions that are progressively and coherently farther away from the pixel centers.  In the left and right panels of Figure~\ref{fig:pk_pdf_bilinear_interp}, we show the resulting power spectra and the PDF of the convergence maps that have been rotated and interpolated according to the above procedure.  

In our ray-tracing scheme, all the rays start from an observed angular position that coincides with the pixel centers, but as these are propagated from plane to plane, their angular positions will be displaced at every step. We therefore expect the overall smoothing effect of bilinear interpolation for each ray and for each lens plane to correspond to an effective smoothing over an intermediate angular offset from the pixel centers. This is exactly what we observe in both panels of Figure~\ref{fig:pk_pdf_bilinear_interp}, where the line referring to ray-tracing with bilinear interpolation lies consistently within the set of lines indicating increasing values of $\Delta \phi$, closely sticking to the line with $\Delta \phi = 0.2 \phi_{\rm{pix}}$.

\bsp
\label{lastpage}
\end{document}